\documentclass{article}

\usepackage[preprint]{neurips_2024}
\usepackage[utf8]{inputenc} %
\usepackage[T1]{fontenc}    %
\usepackage{hyperref}       %
\usepackage{url}            %
\usepackage{booktabs}       %
\usepackage{amsfonts}       %
\usepackage{nicefrac}       %
\usepackage{microtype}      %
\usepackage{xcolor}         %

\title{VoiceShop: A Unified Speech-to-Speech Framework for Identity-Preserving Zero-Shot Voice Editing}

\usepackage{graphicx}
\usepackage{subcaption}
\usepackage{caption}
\usepackage{wrapfig}
\usepackage{multirow}
\usepackage{pifont}
\usepackage{amsmath}
\usepackage{amssymb}
\usepackage{mathtools}
\usepackage{amsthm}
\usepackage{lipsum}
\usepackage{arydshln}
\usepackage[capitalize,noabbrev]{cleveref}

\newcommand{\xmark}{\ding{55}}

\usepackage[textsize=tiny]{todonotes}
\graphicspath{{figures/}{pictures/}{images/}{./}}

\author{
    Philip Anastassiou$^*$ \quad 
    Zhenyu Tang$^*$ \quad 
    Kainan Peng \quad 
    Dongya Jia \quad \\
    \textbf{Jiaxin Li} \quad
    \textbf{Ming Tu} \quad 
    \textbf{Yuping Wang} \quad 
    \textbf{Yuxuan Wang} \quad 
    \textbf{Mingbo Ma} \\
    Data-Speech Team, ByteDance, San Jose, CA, USA \\
    \texttt{\{philipanastassiou,zhenyu.tang\}@bytedance.com}
}

\begin{document}
\maketitle
\def\thefootnote{*}\footnotetext{Equal contribution}\def\thefootnote{\arabic{footnote}}

\begin{abstract}
We present VoiceShop, a novel speech-to-speech framework that can modify multiple attributes of speech, such as age, gender, accent, and speech style, in a single forward pass while preserving the input speaker's timbre. Previous works have been constrained to specialized models that can only edit these attributes individually and suffer from the following pitfalls: the magnitude of the conversion effect is weak, there is no zero-shot capability for out-of-distribution speakers, or the synthesized outputs exhibit undesirable timbre leakage. Our work proposes solutions for each of these issues in a simple modular framework based on a conditional diffusion backbone model with optional normalizing flow-based and sequence-to-sequence speaker attribute-editing modules, whose components can be combined or removed during inference to meet a wide array of tasks without additional model finetuning. Audio samples are available at \url{https://voiceshopai.github.io}.
\end{abstract}

\section{Introduction}
Research efforts in deep generative modeling have historically been restricted to training specialized models for each task within a given domain, limiting their versatility and requiring expertise in the idiosyncrasies of each sub-specialty \citep{baevski2022data2vec, kaiser2017model, pmlr-v139-jaegle21a}. Aided by the steady progress of large-scale pre-training, expanding datasets, and self-supervised learning paradigms, attention has turned to developing task-agnostic ``foundation models'' built on learning universal representations of data, often combining modalities \citep{Li2023MultimodalFM}. The primary advantage of such models is the ability to perform several tasks within a single, unified framework, usually enabled by finetuning on a collection of downstream tasks or the incorporation of multitask learning objectives. In the speech, audio, and music domains, this current has materialized in a series of recent works, such as VALL-E \citep{wang2023neural}, Voicebox \citep{le2023voicebox}, UniAudio \citep{yang2023uniaudio}, VioLA \citep{wang2023viola}, AudioLDM2 \citep{liu2023audioldm}, SpeechT5 \citep{ao2022speecht5}, WavLM \citep{chen2022wavlm}, and MuLan \citep{huang2022mulan}, all of which aim to support multiple audio-related use cases simultaneously, signaling a paradigm shift in the study and use of deep learning systems at large.

Even as the quality of proposed speech synthesis models improves following advances in diffusion \citep{kong2020diffwave} and neural codec modeling \citep{defossez2022high, zeghidour2021soundstream}, disentangled representation learning, whereby attributes such as a speaker's timbre, prosody, age, gender, and accent are extracted and separated from speech signals, remains an open problem \citep{peyser2022disentangled, polyak2021speech, wang2023disentangled}. Disentangling these characteristics is an essential component of any generative model that aims to allow users to flexibly modify certain attributes of their speech while keeping others constant. Due to the non-trivial nature of this task, several works focus on editing a single attribute at a time and frequently require that synthesized speech map to a set of in-domain target speakers seen during training, leaving the development of systems capable of editing multiple attributes at once under-explored. Therefore, the prevailing limitation of existing models is that such fine-grained control of generated speech in a user's own voice is limited.

To this end, we present VoiceShop, a novel speech-to-speech foundation model capable of a wide assortment of zero-shot synthesis tasks. Within a unified framework, VoiceShop is capable of monolingual and cross-lingual voice conversion, identity-preserving many-to-many accent and speech style conversion, and age and gender editing, where all tasks may be performed in a zero-shot setting on arbitrary out-of-domain speakers not seen during training. Furthermore, our framework enables users to perform any combination of these tasks simultaneously in a single forward pass while preserving the original speaker's unedited attributes, i.e., one may edit the age, gender, accent, and speech style of input speech at once. 

Thus, VoiceShop's capabilities go beyond the scope of traditional \emph{voice conversion} (VC), which aims to synthesize speech that retains the linguistic information of an utterance by a source speaker, while applying the timbre of a target speaker. Instead, our work addresses \emph{voice editing} (VE), which we define as the modification of disentangled speech attributes or the creation of new voices solely based on the source speaker without requiring a target speaker. 

This behavior is achieved through the use of a diffusion backbone model, which accepts global speaker embeddings and time-varying content features as conditioning signals, enabling robust zero-shot VC. We additionally train two separate task-specific editing modules based on a continuous normalizing flow model  \citep{rezende2016variational} that operates on the global speaker embedding to achieve age and gender editing and a sequence-to-sequence model \citep{sutskever2014sequence} that operates on the local content features to achieve accent and speech style conversion. Together, these components make up VoiceShop's modular framework and enable its flexible VE capabilities without finetuning. Our main contributions are as follows:
\begin{itemize}
    \item {\bf Multi-attribute speech editing via unified scalable framework:} VoiceShop supports both conventional VC and VE in a single speech-to-speech framework. Editing modules are trained separately from the backbone synthesis module and may be used in a modular plug-and-play fashion, making them scalable to novel editing tasks and datasets without additional finetuning. 
    \item {\bf Zero-shot generalization:} Our model achieves strong zero-shot VC and VE performance on arbitrary unseen speakers, matching or outperforming existing state-of-the-art (SOTA) baselines specialized for different sub-tasks.
    \item {\bf Disentangled attribute control:}  We achieve strong disentanglement of multiple speaker attributes, which in the VE case allows simultaneous identity-preserving transformations of users' voices without modifying their timbre.
\end{itemize}

\section{Related Work}

\begin{figure*}[t]
\centering
\includegraphics[width=\linewidth]{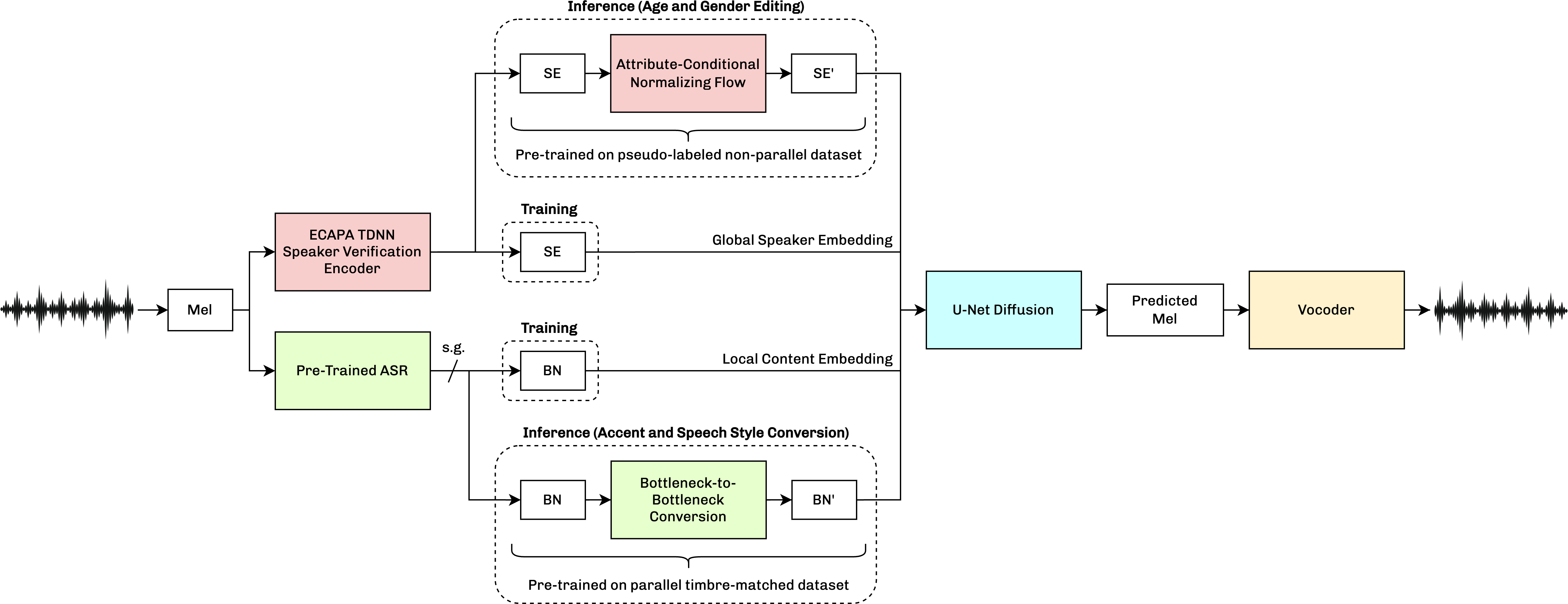}
\caption{{\bf The architecture of VoiceShop:} The overall pipeline follows an analysis-synthesis approach. During the analysis step, the ECAPA-TDNN speaker encoder and pre-trained ASR module respectively decompose input speech into speaker identity represented by a global speaker embedding (SE) and local content embeddings represented by a sequence of bottleneck (BN) features. During the synthesis step, both SE and BN features condition the diffusion backbone model to reconstruct mel-spectrograms of input speech, followed by a vocoder to acquire time-domain waveforms. {\bf Voice editing:} Editing is an optional step in between the analysis and synthesis steps reserved for inference. As highlighted in the two dashed boxes, an attribute-conditional flow module is used to globally edit the speaker embedding (e.g., change age and gender), whereas a BN2BN module is used to edit content embeddings (e.g., change prosody or speech style). These voice editing modules are trained separately from the generative backbone module and are used in a modular plug-and-play manner.}
\label{overview}
\end{figure*}

\paragraph{Zero-Shot Voice Conversion.} While fully supervised VC is considered a canonical problem in the speech domain, AutoVC \citep{qian2019autovc} is among the earliest works to successfully extend VC to the zero-shot setting, demonstrating that promising results could be achieved using a simple autoencoder architecture and carefully designed information bottleneck. Since this work, a variety of methods has been proposed to improve the generalization abilities of zero-shot any-to-any VC systems, such as end-to-end normalizing flow models that maximize the variational lower bound of the marginal log-likelihood of data \citep{casanova2022yourtts, kim2021conditional, li2022freevc, pankov2023dinovits}, unsupervised mutual information-based strategies for speaker-content disentanglement \citep{wang2021vqmivc}, and diffusion-based approaches \citep{choi2023dddmvc, kim2023unitspeech, popov2021diffusion}. While many of these proposals achieve high-quality outputs, the cross-lingual case \citep{casanova2022yourtts}, whereby source and target utterances are spoken in different languages, is less explored. In this work, we show that it is possible to achieve strong zero-shot VC performance using a relatively simple diffusion modeling approach while supporting monolingual and cross-lingual conversion.

\paragraph{Accent and Speech Style Conversion.} \textit{Style transfer} refers to the task of disentangling stylistic and semantic information of samples from a given domain in order to synthesize new samples that impose the style of a reference sample, while retaining the content of another \citep{gatys2015neural}. \textit{Accent conversion} (AC) or \textit{speech style conversion} refines this definition to focus on converting the accent or ``speech style'' of an utterance, i.e., modifying a speaker's pronunciation or prosody, while preserving their spoken content and timbre. Several works have been proposed to achieve accent or speech style conversion, but are often limited by any combination of the following factors. The models may lack zero-shot capabilities, requiring that input timbres are mapped to a set of in-domain target timbres seen during training, altering the identity of the source speaker \citep{zhang2022mix, accentspeech}, or rely on adversarial training strategies, such as domain adaptation via gradient reversal introduced by \citet{ganin2016domainadversarial}, or empirically designed data augmentation methods to encourage disentanglement \citep{badlani2023multilingual, chan2022speechsplit, choi2022nansy++, jia2023zeroshot, li2022crossspeaker, li2022stylettsvc, zhang2023iemotts}. Many frameworks are designed for text-to-speech (TTS), but do not address the speech-to-speech (STS) case \citep{guan2023interpretable, tinchev2023modelling, zhou2023accented}, require ground truth alignments between input and output features acquired through third-party forced alignment tools \citep{ezzerg2022remap, karlapati2022copycat2}, lack explicit control of specific attributes \citep{wang2018style}, or do not support many-to-many conversion, i.e., one must train separate models for each desired target accent, rather than encompassing all conversion paths in a single model \citep{zhao2018accent, zhao2019fac}. To our knowledge, no works have been proposed for cross-lingual AC, whereby accents extracted from speech of one language are transferred to speech of another language.

\paragraph{Age and Gender Editing.} Within the context of VE, NANSY++ \citep{choi2022nansy++} is the most relevant work to support age and gender editing in a zero-shot manner, achieved through a unified framework for synthesizing and manipulating speech signals from analysis features. In their voice designing pipeline NANSY-VOD, the authors employ three normalizing flow networks in a cascaded manner, where the outputs of the earlier models are passed as conditions to subsequent models, to predict $F_0$ statistics, a global timbre embedding, and a fixed number of timbre tokens used to edit the age and gender of output speech. We show that VoiceShop achieves age and gender editing capabilities using a simpler design consisting of one normalizing flow model that allows simultaneous attribute editing. 

\section{VoiceShop}
\subsection{Method Overview}
The architecture of VoiceShop is depicted in \autoref{overview}. The core modules are trained separately and then frozen in subsequent stages, as follows:
\begin{enumerate}
    \item Train an automatic speech or phoneme recognition (ASR/APR) model to extract intermediate feature maps as time-varying content representations of speech.
    \item Jointly train a speaker encoder and diffusion backbone model (optionally including a neural vocoder) conditioned on the time-varying content features and global utterance-level embeddings produced by the speaker encoder to predict mel-spectrograms of speech.
    \item Train individual attribute editing modules to be used during inference for individual or combined multi-attribute editing of the source speaker's voice.
\end{enumerate}
The first two stages listed above consist of large-scale pre-training of the diffusion backbone model to achieve robust zero-shot VC ability. The third stage focuses on separately preparing lightweight modules that respectively operate on the diffusion model's two conditioning signals to modify one or more speech attributes such as age, gender, accent, or speaking style during inference. By tackling VE in this modular approach, we remove the need for transfer learning via finetuning on downstream tasks later on.

\subsection{Large-Scale Pre-Training}
We describe the technical details of three pre-trained models vital to our proposed framework: the \emph{conformer-based ASR model}, the \emph{conditional diffusion backbone model}, and the \emph{vocoder}. During the large-scale pre-training stage of these models, it is crucial to respectively train on large amounts of diverse speech data from several speakers in various recording conditions. Doing so ensures that the models learn sufficiently generalized distributions, which is necessary for zero-shot inference, while indirectly improving the performance of the attribute-editing modules, whose ground truth targets are extracted from these pre-trained models. To fully utilize all available data, which often does not include textual transcriptions, we adopt fully self-supervised training schemes for all aforementioned models, with the exception of the ASR model, whose details we provide in the following sections.

\subsubsection{Conformer-based ASR Model}
\label{sec:asr}
In principle, any parametric model that produces time-varying content representations, whether extracted from models optimized using traditional ASR criteria \citep{baevski2020wav2vec, chan2021speechstew, majumdar2021citrinet} or more recent universal speech frameworks based on vector quantization methods to enable language modeling objectives \citep{wang2023viola, yang2023uniaudio, yang2023universal, zhang2023google}, may be used in our framework. In order to enable cross-lingual synthesis capabilities, we train our own monolingual and bilingual conformer-based ASR models \citep{Gulati_2020} from scratch, referred to as \emph{ASR-EN} and \emph{ASR-EN-CN} respectively, such that the former only transcribes English speech and the latter transcribes both English and Mandarin speech, based on the open-source ESPnet\footnote{\url{https://github.com/espnet/espnet}} \citep{watanabe2018espnet} library. For implementation details, please refer to \S\ref{sec:asr_details}.

\subsubsection{Conditional Diffusion Backbone Model for Zero-Shot Voice Conversion}
\label{sec:backbone}

Denoising diffusion probabilistic models (DDPMs)  \citep{ho2020denoising} are a class of latent variable models that employ two Markovian chains, referred to as the ``forward'' and ``reverse'' processes. In the forward process $q(x_{1:T}|x_0)$, data samples $x_0 \sim q(x_0)$ are iteratively injected with Gaussian noise for time steps $t\in[1,T]$ according to a deterministic variance schedule until corrupted. In the learnable reverse process $p_\theta(x_{0:T})$, a neural network is tasked with approximating $q(x_{t-1}|x_t)$, which is intractable, by predicting and removing noise added in the forward process for each time step $t$, until the original sample is retrieved, thereby learning a mapping between the original data distribution and a tractable prior distribution, such as a standard isotropic Gaussian distribution \citep{yang2024diffusion}.

We propose a conditional diffusion model to predict mel-spectrogram representations of speech, serving as the backbone of our unified framework. Specifically, we consider a reverse process $p_\theta(x_{0:T}|E_S, E_C)$, where $x_0\in \mathbb{R}^{F\times L}$ denotes a mel-spectrogram extracted from raw audio, such that $F$ is the number of frequency bands and $L$ refers to the duration, $S(x_0)=E_S\in \mathbb{R}^{D_S}$ denotes an utterance-level global speaker embedding (i.e., lacking temporal information) produced by speaker encoder $S$, and $C(x_0)=E_C\in \mathbb{R}^{D_C\times\frac{L}{4}}$ denotes a time-varying content embedding produced by a pre-trained ASR or APR model $C$. We train the speaker encoder jointly with the diffusion model from scratch, adopting the same configuration as the ECAPA-TDNN speaker verification model \citep{desplanques2020ecapa} to extract timbre information as a 512-dimensional vector.

As with Moûsai \citep{schneider2023mousai}, we employ a one-dimensional U-Net \citep{ronneberger2015unet} as the architecture of the diffusion model and modify their open-source implementation based on the A-UNet toolkit\footnote{\url{https://github.com/archinetai/audio-diffusion-pytorch}}. We follow their design for each U-Net block with some modifications, the details of which are discussed in \S \ref{sec:diffusion_details}. We also adopt the velocity-based formulation of the diffusion training objective and angular reparametrization of DDIM \citep{song2022denoising} proposed by \citet{salimans2022progressive}, defining our forward process as: 
\begin{equation}
    q(x_t|x_0)=\mathcal{N}(x_t;\alpha_t x_0,\beta_t \epsilon)
\end{equation}
where $\alpha_t \coloneqq \cos{\left(\frac{t\pi}{2}\right)}$, $\beta_t \coloneqq \sin{\left(\frac{t\pi}{2}\right)}$, $x_t = \alpha_t x_0 - \beta_t \epsilon$ denotes noisy data at time step $t\sim\mathcal{U}[0,1]$, and $\epsilon \sim \mathcal{N}(\mathbf{0},\mathbf{I})$. Under this setup, the model minimizes the mean squared error (MSE) between ground truth and predicted velocity terms, as follows:
\begin{align}
\label{eq_diff_loss}
    \mathcal{L}(\theta) \coloneqq{}& \mathbb{E}_{t\sim\mathcal{U}[0,1]} \left[ \|v_t-f_{\theta} (x_t;t,E_S,E_C)\|_2^2 \right] \\
     v_t ={}& \alpha_t \epsilon- \beta_t x_0
\end{align}
where $\theta$ denotes the parameters of the speaker encoder and diffusion model. During inference, we generate samples from noise with a DDIM sampler by applying the following over uniformly sampled time steps $t\in[0,1]$ for $T$ steps: 
\begin{align}
    \hat{v}_t &= f_{\theta} (x_t;t,E_S,E_C)\\
    \hat{x}_0 &= \alpha_t x_t - \beta_t \hat{v}_t \\
    \label{eq_ddim_sampler} \hat{\epsilon}_t &= \beta_t x_t + \alpha_t \hat{v}_t\\ 
    \hat{x}_{t-1} &= \alpha_{t-1} \hat{x}_0 + \beta_{t-1} \hat{\epsilon}_t
\end{align}
When deciding which content features to use for conditioning the diffusion backbone model, we consider observations made by \citet{yang2023accent}, who find that the selection of which layer to extract content representations from pre-trained ASR or APR models has a measurable impact on the magnitude of various information sources encoded in the latent sequence. Our work validates this finding, and we note that the activation maps of shallower layers of such models, such as the 10$^\text{th}$ layer, contain far greater amounts of prosodic and accent information compared to those of deeper layers, such as the 18$^\text{th}$ layer. Intuitively, this phenomenon may be explained by a layer's proximity to the final loss calculation, such that deeper layers are in some sense ``closer'' to pure textual transcriptions of input speech, while shallower layers still contain significant amounts of timbre and prosody leakage desirable for accent and speech style conversion. 

For this reason, we train four versions of the diffusion backbone model, two of which accept the outputs of the 10$^\text{th}$ and 18$^\text{th}$ layers of the monolingual \emph{ASR-EN} model, respectively denoted \emph{VS-EN-L10} and \emph{VS-EN-L18}, and an additional two which accept those of the 10$^\text{th}$ and 18$^\text{th}$ layers of the bilingual \emph{ASR-EN-CN} model, respectively denoted \emph{VS-EN-CN-L10} and \emph{VS-EN-CN-L18}. The training datasets are listed in \autoref{tab:datasets} and all backbone model configurations are summarized in \autoref{diff-configs}.  

\begin{table}
\parbox{.54\linewidth}{
\centering
\caption{Datasets used to train diffusion backbone models. Monolingual models only use English data, whereas bilingual models use all listed data.}
\label{tab:datasets}
\begin{small}
\setlength{\tabcolsep}{2pt}
\begin{tabular}{lccc}
\toprule
\textbf{Corpus} & \textbf{Language} & \textbf{Hours} \\
\midrule
Common Voice 13.0 & English & 3,209\\
LibriTTS & English & 585 \\
L2-ARCTIC & English & 20 \\
Proprietary ASR/TTS corpus & English & 1,400 \\
AISHELL-3 & Mandarin & 85  \\
Proprietary TTS corpus & Mandarin & 84 \\
\bottomrule
\end{tabular}
\end{small}
}
\hfill
\parbox{.43\linewidth}{
\centering
\caption{Training configurations of each diffusion backbone model. Layer numbers (10 and 18) are contained in model names.}
\label{diff-configs}
\begin{small}
\setlength{\tabcolsep}{3pt}
\begin{tabular}{llc}
\toprule
\textbf{Backbone Model} & \textbf{ASR Model}& \textbf{Bilingual} \\
\midrule
VS-EN-L10 & ASR-EN & \xmark \\
VS-EN-L18 & ASR-EN & \xmark \\
VS-EN-CN-L10 & ASR-EN-CN & \checkmark  \\
VS-EN-CN-L18 & ASR-EN-CN & \checkmark \\
\bottomrule
\end{tabular}
\end{small}
}
\end{table}

We train all diffusion backbone models using the AdamW optimizer \citep{loshchilov2019decoupled} on 4 A100 GPUs with a learning rate of $1 \times 10^{-4}$ and batch size of 88 samples for 250K iterations. We set a learning rate decay of 0.85 applied every 20 epochs. To leverage large amounts of speech data, we train with unlabeled audio in a self-supervised manner. After convergence, the diffusion backbone model becomes capable of robust zero-shot VC.

\subsubsection{Neural Mel-Spectrogram Vocoder}

As the final step of our inference pipeline, we convert the predicted mel-spectrogram of the diffusion model from its time-frequency representation to a time-domain audio signal using a neural vocoder. Like the ASR model, while any SOTA vocoder may be used for this task, we choose to train our own model based on HiFi-GAN \citep{kong2020hifigan} from scratch for improved robustness and uncompromised audio quality when scaling up. For implementation details, please refer to \S\ref{sec:vocoder_details}.

\subsection{Task-Specific Voice Editing Modules}
Rather than finetuning the backbone model obtained in \S\ref{sec:backbone}, we develop individual voice attribute editing modules as flexible, modular plug-ins to the generative pipeline. 

\subsubsection{Attribute-Conditional Normalizing Flow for Age and Gender Editing}
\label{sec:attribute-cnf}
We observe that the speaker encoder jointly trained with our diffusion backbone model achieves strong speech attribute disentanglement, indicating that many attributes like age and gender are encoded into the global 512-dimensional speaker embedding vector. Therefore, the manipulation of these attributes can be seen as re-sampling from the learned latent space of speaker embeddings. To achieve fully controllable generation and editing of specific speaker attributes while leaving other attributes unaffected, we take inspiration from StyleFlow \citep{abdal2021styleflow} in the image editing domain and employ a similar attribute-conditional continuous normalizing flow (CNF) that operates on this latent space. 

\begin{figure}[t]
\vskip 0.1in
\begin{center}
\centerline{\includegraphics[width=0.8\columnwidth]{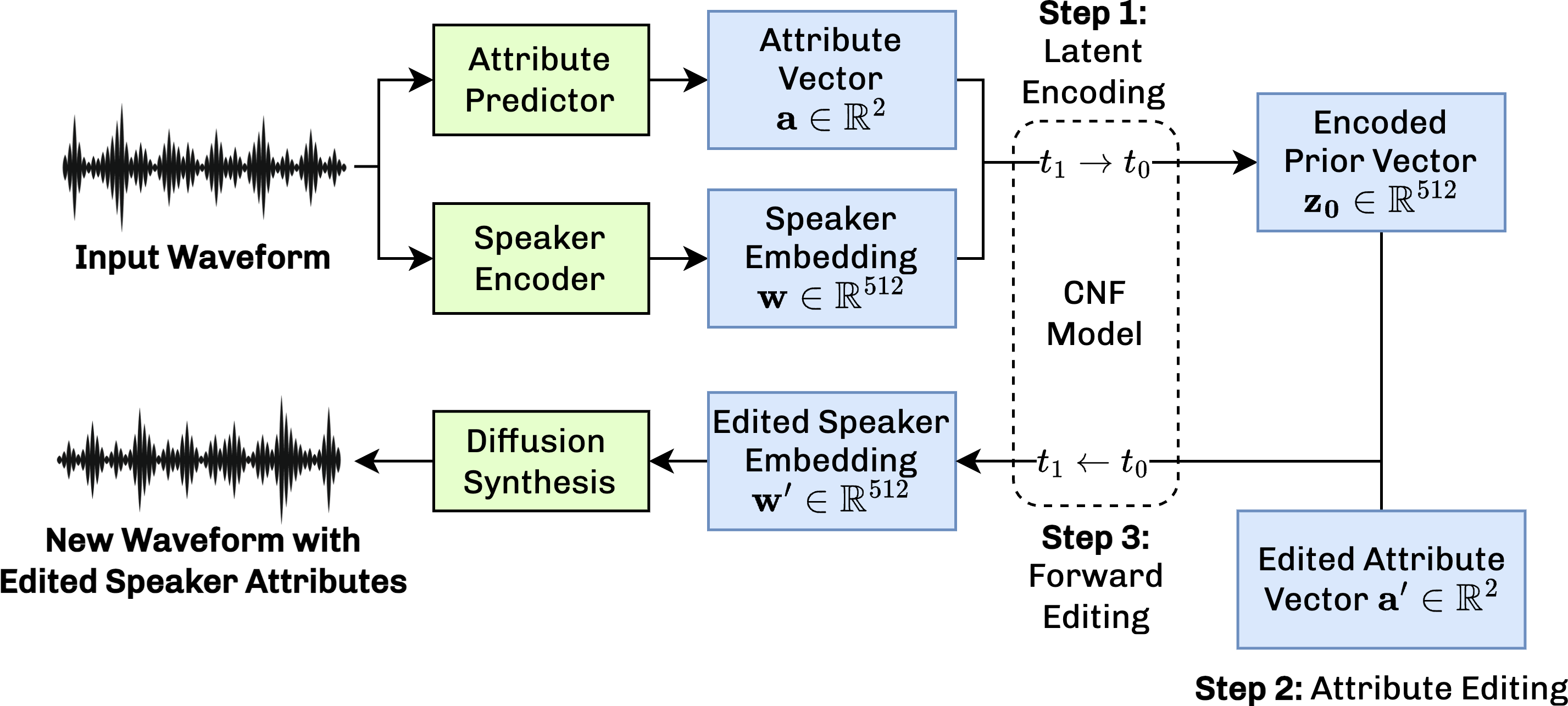}}
\caption{{\bf Attribute conditional flow editing module}: Starting with a speaker's voice sample, we use a pre-trained attribute predictor to obtain age and gender labels, which we denote as the original attribute $\mathbf{a}$, and use the speaker encoder jointly trained with the diffusion backbone to extract the speaker embedding $\mathbf{w}$. The three steps of editing during inference proceed as follows: 1. An ODE solver utilizes the pre-trained CNF model which is conditioned on $\mathbf{a}$ and $t$ to reverse integrate $\mathbf{w}$ from $t_1$ to $t_0$ into $\mathbf{z_0}$, which is the encoded latent in the prior space. 2. Modify any or all attributes of the original speaker to obtain the new attribute vector $\mathbf{a'}$. 3. Use the ODE solver again for forward integration from $t_0$ to $t_1$ using the CNF model with $\mathbf{z_0}$ conditioned on $\mathbf{a'}$. The output is a new speaker embedding $\mathbf{w'}$, which embeds the edited attributes. When using $\mathbf{w'}$ with our diffusion backbone model, the generated voice should retain the unedited attributes in the original input voice (i.e., editing gender should not affect age and vice versa).}
\label{flow-modeling}
\end{center}
\vskip -0.2in
\end{figure}

Continuous normalizing flows utilize a neural ordinary differential equation (ODE) formulation \citep{chen2018neural} to model the bidirectional mapping of two distributions:
\begin{align}
\label{eq:cnf_integral}
\frac{dz}{dt}&=\phi_{ \theta}(z(t), t)\\
z\left(t_1\right)&=z\left(t_0\right)+\int_{t_0}^{t_1} \phi_{ \theta}(z(t), t) d t
\end{align}
where $t$ is the (virtual) time variable, $z(t)$ is the variable of a given distribution, and $\phi_{\theta}$ is an arbitrary neural network that generates outputs that have the same dimensionality as the inputs, parameterized by $\theta$. 
We denote our target distribution of speaker embeddings as $\mathbf{z}\left(t_1\right)=\mathbf{w}\in \mathbb{R}^{512}$ in $W$ space, and variables from a prior distribution as $\mathbf{z}\left(t_0\right)\in\mathbb{R}^{512}$ in $Z$ space, which we conveniently set to a zero-mean multi-dimensional Gaussian distribution with identity variance $\mathcal{N}(\mathbf{0},\mathbf{I})$. By applying the change of variable rule, we can formulate the change in the log density as:
\begin{equation}
\label{eq:cnf_change_log}
\log p\left(\mathbf{w}\right)=\log p\left(\mathbf{z}\left(t_0\right)\right)-\int_{t_0}^{t_1} \operatorname{Tr}\left(\frac{\partial \phi}{\partial \mathbf{z}(t)}\right) d t
\end{equation}
The training objective of the CNF would be to maximize the likelihood of the data $\mathbf{w}$. In our attribute conditional scenario, given the speaker attributes vector $\mathbf{a}$ (e.g., for age and gender, $\mathbf{a}\in \mathbb{R}^2$) that is associated with sample $\mathbf{w}$, we update the ODE network to be $\phi_{ \theta}(z(t), \mathbf{a}, t)$, which is conditioned on both $t$ and $\mathbf{a}$. Consequently, our new objective becomes $\max_{\theta}\sum_{\mathbf{w},\mathbf{a}}\log p(\mathbf{w}|\mathbf{a},t)$. During training, we also employ trajectory polynomial regularization \citep{huang2021accelerating}, which we find stabilizes the training. Note that there is no attribute editing during the training process. A well-trained CNF model behaves as a bijective mapping between the data distribution and prior distribution with the help of an ODE solver. In our experiments, we use the \emph{dopri5} ODE solver \citep{hairer2008solving} and adopt the same CNF implementation from StyleFlow\footnote{\url{https://github.com/RameenAbdal/StyleFlow}}.
We only use one CNF block with hidden dimension 512 rather than stacking multiple blocks, which leads to lower validation loss. Our lightweight CNF model has 0.5M learnable parameters and is trained using Adam \citep{kingma2014adam} on a single V100 GPU for 24H. The attribute editing procedure at inference time is depicted in \autoref{flow-modeling}. 

\paragraph{Attributes Dataset.} The attribute labels for training our attribute-conditional CNF are obtained from the CommonVoice 13.0 dataset \citep{ardila2019common}, which contains multiple audio recordings of more than 51K human-validated anonymous speakers, totaling 2,429 hours. 
Among these validated speakers, about 20K speakers provide age (from twenties to nineties) and gender labels (male or female). To fully utilize the dataset, we first train a naive ECAPA-TDNN-based age and gender prediction model by appending a projection layer that predicts age as a numeric value and gender as a logit to its original output layer. We combine mean absolute error (MAE) and cross-entropy (CE) losses for age and gender prediction respectively. This prediction model serves two purposes:
\begin{enumerate}
    \item We use it to weakly label the remainder of the CommonVoice English dataset for CNF training.
    \item For gender labels, we use the predicted logits instead of binary labels, which facilitates continuous gender editing similar to age labels. 
\end{enumerate}
Without extensive parameter tuning, we obtain a model with 4.39 mean absolute age error and 99.1$\%$ gender accuracy on a holdout test consisting of 10$\%$ of the labeled speakers, after training on 8 V100 GPUs for 48 hours using the AdamW optimizer \citep{loshchilov2019decoupled} with a learning rate of $1 \times 10^{-3}$ and weight decay of $1 \times 10^{-6}$. 
We then use the speaker encoder to extract utterance-level speaker embeddings and use the predicted age and gender labels for these 51K speakers to train CNF. 

\begin{figure}[t]
\vskip 0.2in
\begin{center}
\centerline{\includegraphics[width=0.8\columnwidth]{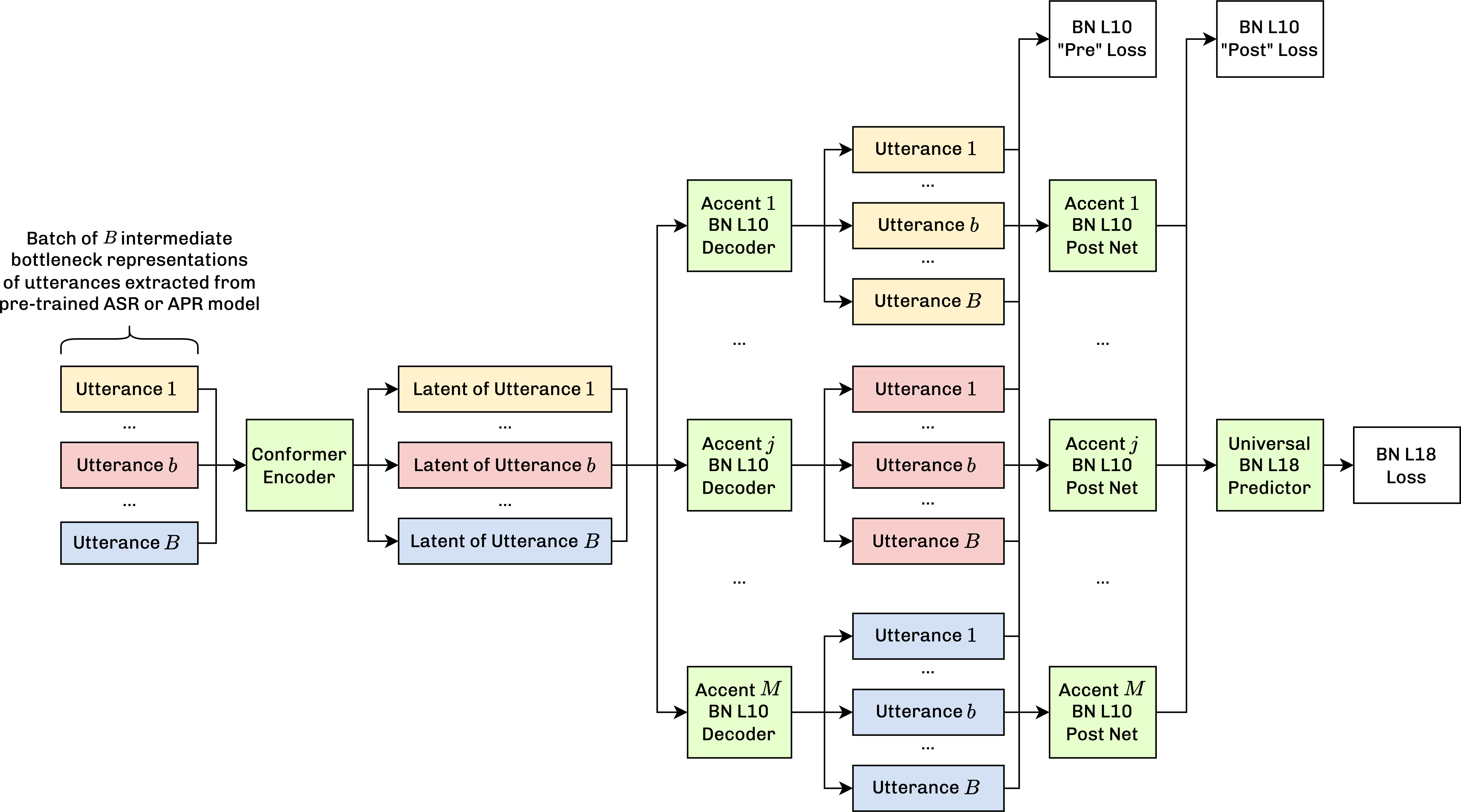}}
\caption{{\bf Bottleneck-to-bottleneck (BN2BN) modeling:} Our BN2BN design maps the time-varying content features of utterances from an arbitrary number of source accents to those of an arbitrary number of target accents in a single model using a multi-decoder architecture.}
\label{bn2bn}
\end{center}
\vskip -0.2in
\end{figure}

\subsubsection{Bottleneck-to-Bottleneck (BN2BN) Modeling for Many-to-Many Accent and Speech Style Conversion}

We propose a bottleneck-to-bottleneck (BN2BN) model capable of many-to-many accent and speech style conversion by mapping the time-varying ``bottleneck'' content features of utterances from an arbitrary number of source accents to those of an arbitrary number of target accents in a single model using a multi-decoder architecture based on encoder-decoder sequence-to-sequence modeling with cross attention, effectively reducing the accent conversion task to a machine translation problem. Beyond typical accent conversion, the BN2BN model is also capable of generalized \emph{speech style} transfer. There are no specific requirements for what constitutes a ``speech style,'' which may be as broad as emotional speech or the speaking styles of iconic personalities from popular culture.

To train in this fashion, we propose a simple method for augmenting non-parallel speech datasets into parallel multi-speaker, multi-accent ``timbre-matched'' datasets by leveraging TTS and VC modeling, only requiring a text corpus at minimum, which we describe in \S\ref{sec:timbre-matched}. In our experiments, we use publicly available TTS models provided by the Microsoft Azure AI platform\footnote{\url{https://azure.microsoft.com/en-us/products/ai-services/text-to-speech}} to synthesize speech in a variety of accents in English and Mandarin for training.

Our approach is rooted in the observation that these latent sequences not only encode the semantics of a speech signal (i.e., \emph{what} is said), but also pronunciation and prosody (i.e., \emph{how} it is said). During training, each batch $X\in\mathbb{R}^{B\times T \times D_1}$ consists of the local content features of utterances, such that the source accent, content, and speaker of inputs are sampled uniformly at random. The utterances are processed by a universal encoder $E$ that learns accent-agnostic latent representations $Z=E(X)\in\mathbb{R}^{B\times T \times D_2}$, which are passed as input to autoregressive domain-specific decoders $D_{j}$ and post nets $P_{j}$ for each supported target accent $j\in [1,M]$ to reconstruct the content features in the respective target accents regardless of the source accent, i.e., $P_{j}(D_{j}(Z))=\hat X_{j}\in\mathbb{R}^{B\times T \times D_1}$, as depicted in \autoref{bn2bn}.

We use the 10$^\text{th}$ layer activation maps from our \emph{ASR-EN} and \emph{ASR-EN-CN} models to achieve AC. We optionally train a final conformer-based predictor module which generates the corresponding 18$^\text{th}$ layer activation maps conditioned on predicted 10$^\text{th}$ layer representations to enable combined multi-attribute conversion alongside the flow-based age and gender editing module, which operates most successfully on speaker embeddings produced by our 18$^\text{th}$ layer-based diffusion backbone models, \emph{VS-EN-L18} and \emph{VS-EN-CN-L18}. We find that this is due to the abundant timbre and prosody leakage of shallow content representations, which proves to be beneficial for AC, but in turn limits the editing capabilities of the speaker embedding. In the absence of such information when training on sparse content features, the speaker embedding of our backbone model necessarily assumes more control of various speech attributes, which is preferable for age and gender editing.

We adopt a conformer encoder and LSTM decoders with the same configuration proposed in Tacotron 2 \citep{shen2018natural}, using additive energy-based Dynamic Convolution Attention (DCA) \citep{battenberg2020location} as the cross-attention mechanism between the encoder and decoders. We jointly train ``stop'' gate projection layers for each decoder using binary CE, which produces a scalar between 0 and 1 indicating when to stop generation ($\mathcal{L}_\text{Gate}$). We apply three MAE reconstruction loss terms to the output of the LSTM decoders ($\mathcal{L}_\text{Pre L10}$), post nets ($\mathcal{L}_\text{Post L10}$), and universal predictor ($\mathcal{L}_\text{L18}$). The final loss is formulated as:
\begin{equation}
    \mathcal{L}_\text{BN2BN} = \mathcal{L}_\text{Pre L10} + \mathcal{L}_\text{Post L10} + \mathcal{L}_\text{L18} + \mathcal{L}_\text{Gate}
\end{equation}
We train each BN2BN model using the AdamW optimizer \citep{loshchilov2019decoupled} on 8 A100 GPUs with a batch size of 32 samples and an initial learning rate of $1 \times 10^{-4}$, to which we apply a decay rate of 0.85 every 100 epochs. The rate of convergence typically relies on the size of the dataset and varies between approximately 75 to 200 epochs.

\section{Experiments and Analysis}

We demonstrate the versatile capabilities of VoiceShop on various synthesis-related tasks, specifying each model configuration used.\footnote{Audio samples are available at \url{https://voiceshopai.github.io}.} For all generated samples, we use 5 time steps in the reverse diffusion process and replace \autoref{eq_ddim_sampler} with $\epsilon_t \sim \mathcal{N}(\mathbf{0},\mathbf{I})$ in our DDIM sampler, which we empirically find causes no perceptual difference in output quality. 

\paragraph{Evaluation Metrics.} We evaluate the performance of each mentioned synthesis task using a variety of subjective and objective metrics, defined as follows.

For subjective evaluation, we conduct Mean Opinion Score (MOS) and Comparative Mean Opinion Score (CMOS) studies with anonymous human participants who are tasked with judging the performance of VoiceShop on pre-defined metrics, such as perceived speaker similarity, conversion strength, and naturalness. For MOS studies, participants rank individual samples on a scale between 1 and 5, where 1 indicates poor performance and 5 indicates strong performance, whereas CMOS studies ask participants to consider pairs of samples labeled A and B and rank their preference on a scale between -3 to 3, where -3 indicates a strong preference for sample A, 3 indicates a strong preference for sample B, and 0 indicates no preference.

To objectively evaluate the speaker similarity between ground truth input and synthesized output pairs at a large scale, we use the Automatic Speaker Verification (ASV) metric, whereby we compute the cosine similarity of fixed-length embeddings extracted from a pre-trained speaker verification model. Therefore, the ASV metric falls within the range of -1 and 1, where greater values indicate higher similarity within the learned latent space of the speaker verification model, which is assumed to effectively discriminate timbre. We follow VALL-E \citep{wang2023neural} and Voicebox \citep{le2023voicebox} and adopt WavLM-TDNN\footnote{\url{https://github.com/microsoft/UniSpeech/tree/main/downstreams/speaker_verification}} \citep{chen2022wavlm} to extract speaker embeddings for this purpose. Additional task-specific metrics are defined in subsequent sections as needed.

For tasks where no obvious third-party baselines exist for direct comparisons, such as style conversion and cross-lingual AC, we report scores of intentionally incorrect pairs (i.e., for ASV, we compute the similarity of non-matching speaker pairs). When placed alongside the scores of ground truth inputs, doing so establishes upper and lower bounds against which we can more meaningfully understand the values produced by VoiceShop's outputs. For all applicable metrics, we summarize results with 95\% confidence intervals. 

\subsection{Zero-Shot Voice Conversion} 
We demonstrate VoiceShop's performance on monolingual and cross-lingual zero-shot VC, which refers to the task of modifying the timbre of a spoken utterance while preserving its content for arbitrary speakers not seen during training. In the monolingual case, both the source content and target timbre are spoken in the same language, whereas in the cross-lingual case, they are from different languages (e.g., apply the timbre of a Mandarin utterance to English content). Since the global speaker embedding used in our diffusion backbone model collapses temporal information, we additionally find that conversion can be achieved for out-of-domain languages not seen by the diffusion backbone model during training, enabling anyone to speak fluent English and Mandarin in their own voice regardless of their native language.

\paragraph{Subjective MOS Evaluation.} We evaluate our framework on zero-shot VC against YourTTS \citep{casanova2022yourtts} and DiffVC \citep{popov2021diffusion}, two recent works regarded as SOTA in this task. Following the experimental setups of these works, we use the VCTK corpus \citep{veaux2016superseded} as the test set. We identify 8 VCTK speakers (4 males and 4 females) that are held out by both works, while our model is not trained on any VCTK data. We select one speech sample for each speaker as their reference, then each VC model will convert the voice of each speaker to the voice of all other speakers, yielding 56 samples per model. As ground truth for all test subsets, we randomly select 7 more audios for each of the test speakers, creating another 56 samples. 

In preparing test samples, we directly curated samples from YourTTS's official demo page\footnote{\url{https://edresson.github.io/YourTTS/}}, and generated the same set of samples using DiffVC's official implementation\footnote{\url{https://github.com/trinhtuanvubk/Diff-VC}} with their best performing \emph{Diff-LibriTTS-wodyn} checkpoint for a stronger baseline. For the purpose of this study, we instead use our \emph{VS-EN-L18} backbone model to perform inference, as it is restricted to English speech. All test samples are downsampled to 16kHz to match the test condition of YourTTS, and then normalized for loudness. 

\begin{wraptable}{R}{6.2cm}
\caption{Results of subjective MOS evaluation of zero-shot VC for \emph{speaker similarity} (sMOS) and \emph{naturalness} (nMOS).}
\label{vc_mos}
\begin{center}
\begin{small}
\begin{tabular}{lcc}
\toprule
\textbf{Model}  & \textbf{sMOS ($\uparrow$)} & \textbf{nMOS ($\uparrow$)} \\
\midrule
YourTTS & 2.89$\pm$0.07 & 3.04$\pm$0.07 \\
DiffVC & 3.14$\pm$0.07 & 3.42$\pm$0.07 \\
VoiceShop (Ours) & \textbf{3.76$\pm$0.07} & \textbf{3.56$\pm$0.06}\\
\midrule
Ground Truth & 4.24$\pm$0.06 & 4.24$\pm$0.05 \\
\bottomrule
\end{tabular}
\end{small}
\end{center}
\end{wraptable}

We conduct a MOS study with third-party vendors, recruiting 20 native English speakers (12 males, 8 females, mean age 32.0 years with a standard deviation of 6.5 years) for the study, in which each voice-converted sample or ground truth sample is presented together with the reference sample. We ask participants to rate the sample's speech naturalness independently, then rate the sample's speaker similarity with respect to the reference sample, disregarding its naturalness. We adopt a five-point Likert scale for both questions. 
We summarize the MOS study results in \autoref{vc_mos}. In both dimensions, our VoiceShop model outperforms other baseline models by clear margins. 

\subsection{Zero-Shot Identity-Preserving Many-to-Many Accent Conversion} 

\begin{figure}[t]
\vskip 0.2in
\begin{center}
\centerline{\includegraphics[width=0.7\columnwidth]{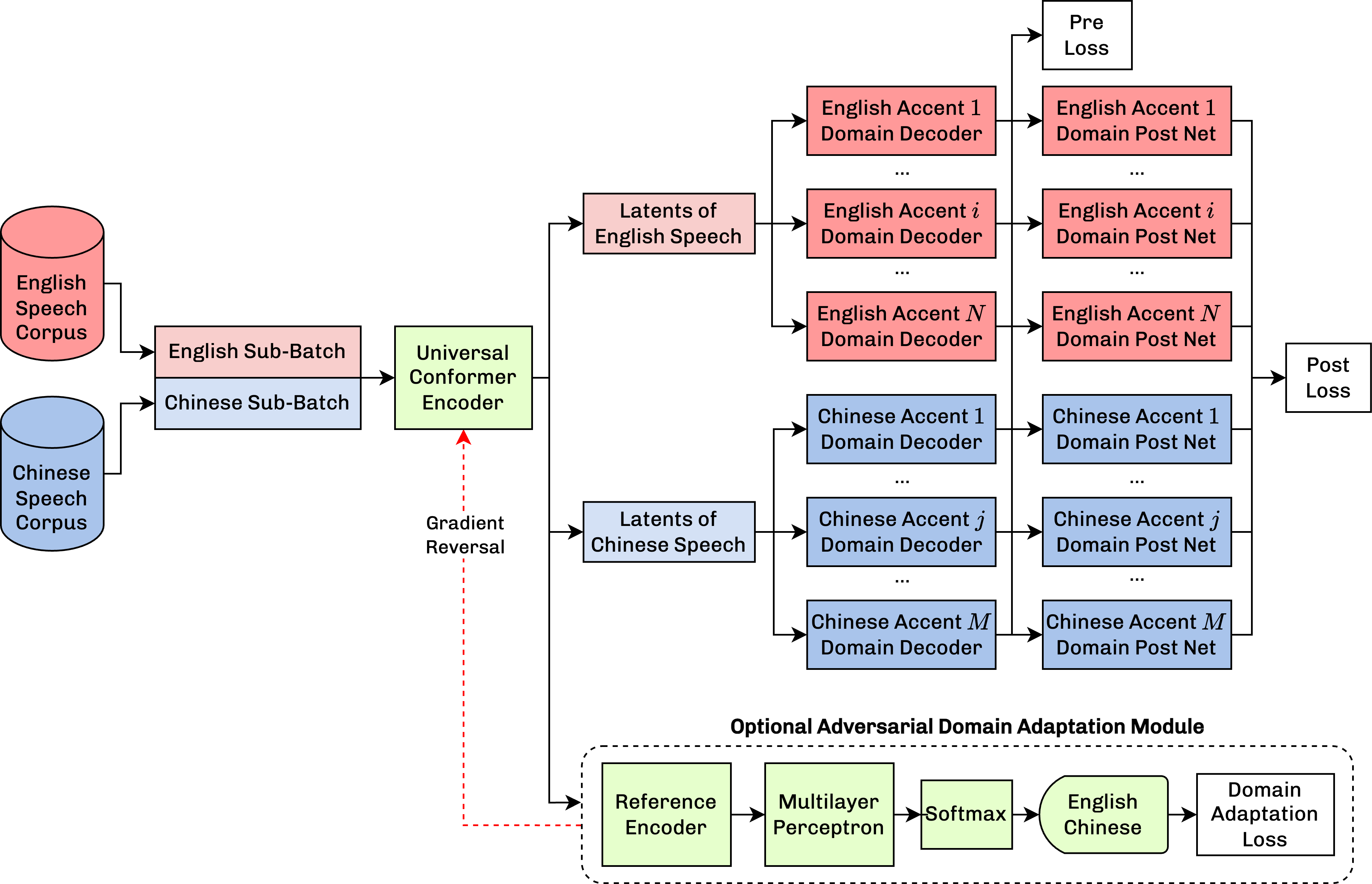}}
\caption{Training configuration of cross-lingual AC using BN2BN modeling, featuring adversarial domain adaptation via gradient reversal to promote language-agnostic content representations in the learned latent space of the universal encoder. We use the same reference encoder proposed in \citet{wang2018style}.}
\label{xlingual-bn2bn}
\end{center}
\end{figure}

We demonstrate VoiceShop's capabilities on monolingual and cross-lingual identity-preserving many-to-many AC in English and Mandarin, which refers to the task of modifying the accent of an utterance while preserving the original speaker's timbre for arbitrary speakers not seen during training. In the monolingual case, AC occurs within the same language, whereas in the cross-lingual case, we convert to accents of a different language in the absence of parallel data (e.g., apply a British accent to Mandarin speech only using British-accented English speech or apply a Sichuan accent to English speech only using Sichuan-accented Mandarin speech, as recordings of British-accented Mandarin speech or Sichuan-accented English speech are not available). We depict modifications to our BN2BN design used only for cross-lingual conversion in \autoref{xlingual-bn2bn}, whereby we add a gradient reversal module to promote language-agnostic representations, setting $\lambda=-1$. In both cases, AC is achieved in a many-to-many manner, i.e., a user can convert an arbitrary number of source accents to an arbitrary number of target accents in one BN2BN model. All samples are generated using our \emph{VS-EN-CN-L10} backbone model. To our knowledge, VoiceShop is the first framework that achieves cross-lingual AC.

\paragraph{Subjective CMOS Evaluation.} We evaluate our model against \citet{jin2023voicepreserving}, a recent work which also achieves zero-shot identity-preserving many-to-many accent conversion, curating 16kHz samples directly from their official demo page\footnote{\url{https://accent-conversion.github.io/}}. As our diffusion backbone model requires 24kHz input, we use the official pre-trained checkpoint of \emph{AudioSR-Speech}\footnote{\url{https://github.com/haoheliu/versatile_audio_super_resolution}}, a versatile audio super-resolution model, to predict the corresponding high-fidelity 48kHz speech, which we then downsample to 24kHz prior to performing inference with our framework. Since upsampling the 16kHz audio is lossy and cannot recover spectral information between 8 to 12kHz in the frequency domain, it is necessary to use a generative model to properly extend the bandwidth. After inference, all samples are downsampled to 16kHz for a fair comparison.

\begin{wraptable}{R}{7.8cm}
\caption{Results of subjective CMOS evaluation of monolingual AC for \emph{accent strength} (aCMOS) and \emph{speaker similarity} (sCMOS).}
\label{ac_cmos}
\begin{center}
\begin{tabular}{ccc}
\toprule
\textbf{Model} & \textbf{aCMOS ($\uparrow$)} & \textbf{sCMOS ($\uparrow$)} \\
\midrule
\multirow{2}{*}{\shortstack{VoiceShop (Ours) vs.\\ \citet{jin2023voicepreserving}}} & \multirow{2}{*}{\shortstack{\textbf{0.89$\pm$0.15}}} & \multirow{2}{*}{\shortstack{0.11$\pm$0.13}} \\ \\
\bottomrule
\end{tabular}
\end{center}
\end{wraptable}

We conduct two CMOS studies to respectively measure the accent strength (aCMOS) and speaker similarity (sCMOS) of our systems using 30 pairs of samples (specifically, we evaluate on their British-to-American, British-to-Indian, and Indian-to-American subsets). Both studies recruit 20 anonymous native English speakers balanced across gender using the Prolific crowd-sourcing platform\footnote{\url{https://www.prolific.com/}}, where participants are asked to rate pairs of samples on a scale from -3 to 3, where -3 indicates a strong preference for the baseline model, 3 indicates a strong preference for VoiceShop, and 0 indicates no preference. Each comparison is made against provided reference clips to standardize the comparisons between listeners, who may have different expectations of the target accents' characteristics. The results are summarized in \autoref{ac_cmos}. We find that participants prefer the accent strength of our system by a clear margin while maintaining relatively equal performance on speaker similarity. 

\paragraph{Objective Evaluation.} We conduct additional objective evaluations to judge VoiceShop's monolingual and cross-lingual AC capabilities. Specifically, we train three accent classifiers consisting of an ECAPA-TDNN encoder \citep{desplanques2020ecapa} and multilayer perceptron on the same datasets used to train our BN2BN models for monolingual English AC, monolingual Mandarin AC, and cross-lingual AC. All classifiers are trained using AdamW \citep{loshchilov2019decoupled} for 24H on 8 V100 GPUs with a batch size of 64, a learning rate of $1 \times 10^{-4}$, and decay rate of 0.85 applied every 100 epochs, achieving validation accuracies of 99.3\% in English, 99.8\% in Mandarin, and 99.9\% in the cross-lingual case on their respective holdout sets.

\begin{wraptable}{R}{7cm}
\caption{Results of monolingual and cross-lingual accent conversion objective evaluation, measuring \emph{speaker similarity}. We denote the number of samples used to calculate each value in parentheses.}
\label{ac_asv}
\begin{center}
\begin{small}
\begin{tabular}{lc}
\toprule
\textbf{Conversion Type} & \textbf{ASV ($\uparrow$)} \\
\midrule
English Monolingual & 0.508$\pm$0.006 (2,000) \\
Mandarin Monolingual & 0.809$\pm$0.003 (1,280) \\
Cross-Lingual & 0.625$\pm$0.010 (1,280) \\
\cmidrule{1-2}
Overall & \textbf{0.625$\pm$0.005} (4,560) \\
\midrule
Same Speaker & 0.925$\pm$0.001 (1,900) \\ 
Non-Matching Speaker & 0.175$\pm$0.005 (3,420) \\
\bottomrule
\end{tabular}
\end{small}
\end{center}
\end{wraptable}

For each conversion type, we synthesize input speech in a variety of accents using the Microsoft Azure TTS service and convert each source accent to each target accent using our models. We use the learned latent space of each classifier to investigate the accent similarity of samples generated by our system, first computing the centroids of embeddings extracted from the last layer of each classifier using reference samples for each target accent. We then extract embeddings of samples to calculate the cosine similarity against the averaged reference embeddings for three categories: a separate set of ground truth samples, the accent-converted outputs of our models, and non-matching accents to establish a performance lower bound. We report overall similarities in \autoref{accent_cosine_similarity} and provide a detailed breakdown of accent-wise similarities in \autoref{accent_cosine_similarity_by_accent}.

\begin{table}[tb]
\caption{Results of monolingual and cross-lingual accent conversion objective evaluation, measuring \emph{conversion strength}. For each target accent, we provide the averaged cosine similarity of embeddings extracted from accent classifiers.}
\label{accent_cosine_similarity}
\begin{center}
\begin{tabular}{lccc}
\toprule
\textbf{Conversion Type} & \textbf{Ground Truth} & \textbf{Model Output} & \textbf{Non-Matching Accent} \\
\midrule
English Monolingual & 0.996$\pm$0.001 & \textbf{0.798$\pm$0.016} & 0.018$\pm$0.004 \\
\cmidrule{1-4}
Mandarin Monolingual & 0.987$\pm$0.003 & \textbf{0.915$\pm$0.011} & 0.621$\pm$0.018 \\
\cmidrule{1-4}
Cross-Lingual & 0.984$\pm$0.002 & \textbf{0.828$\pm$0.015} & 0.550$\pm$0.018 \\
\bottomrule
\end{tabular}
\end{center}
\end{table}

\begin{figure}[t]
\centering
\begin{subfigure}{0.6\linewidth}
\includegraphics[width=\linewidth]{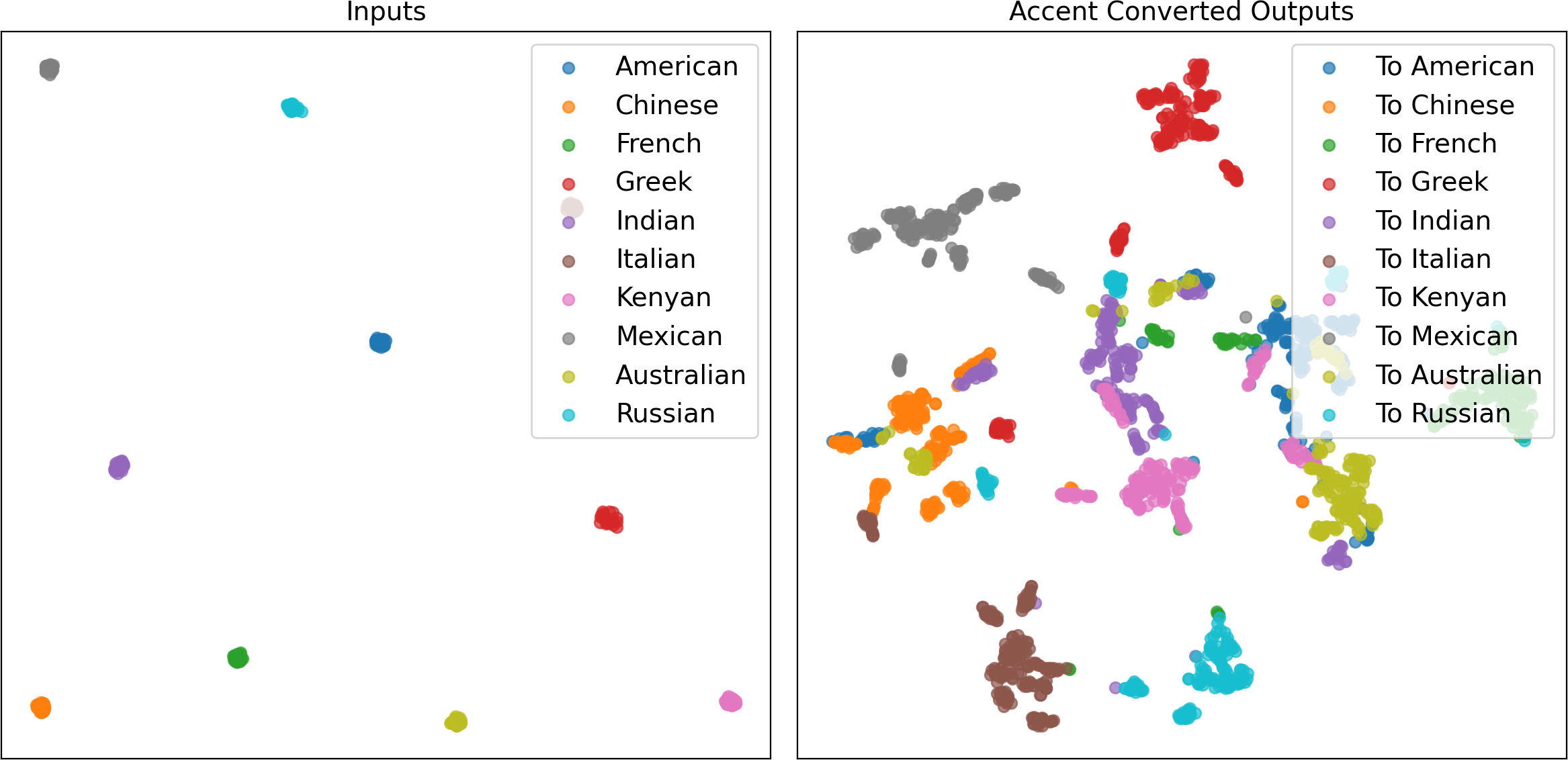}
\caption{Monolingual English accent conversion}
\end{subfigure}
\begin{subfigure}{0.6\linewidth}
\includegraphics[width=\linewidth]{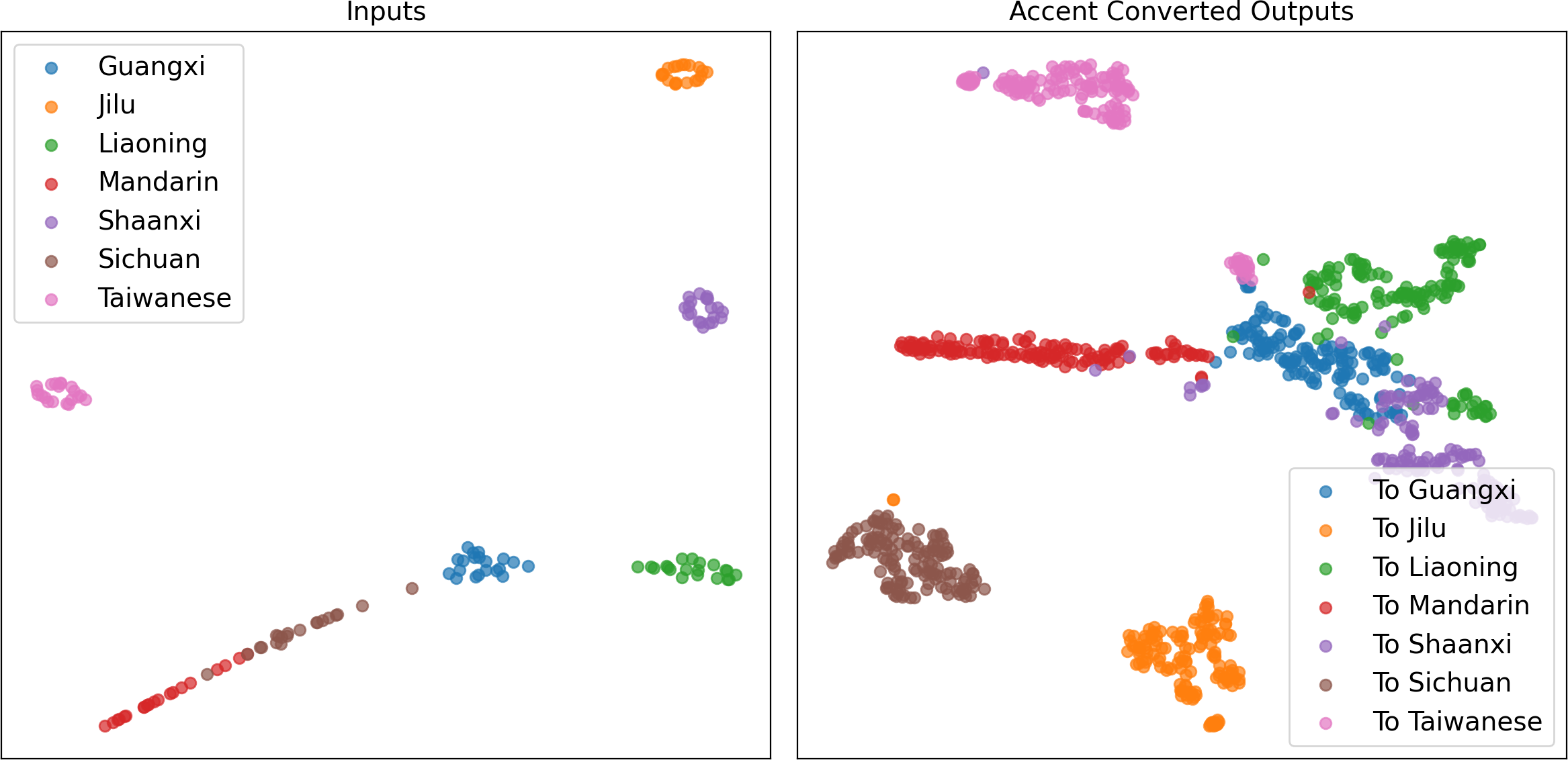}
\caption{Monolingual Mandarin accent conversion}
\end{subfigure}
\begin{subfigure}{0.6\linewidth}
\includegraphics[width=\linewidth]{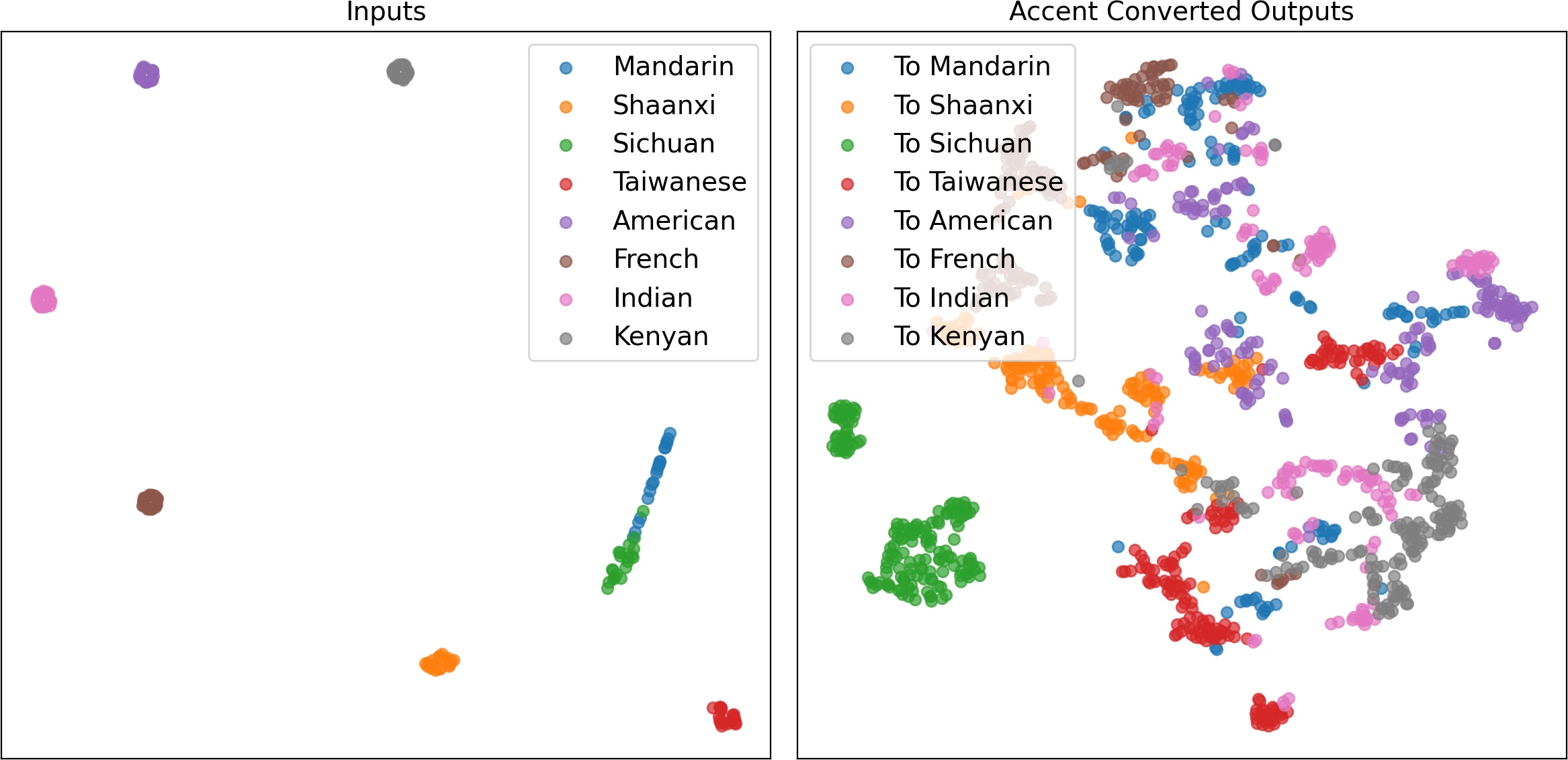}
\caption{Cross-lingual accent conversion}
\end{subfigure}
\caption{\textbf{Visualizing accent transfer:} By using the latent space of accent classifiers, we observe that input speech is clustered by source accent and that accent-converted speech predicted by our BN2BN models largely preserve these structures according to their target accents.}
\label{fig:tsne_plots}
\vspace{0.1cm}
\end{figure}

To visualize these results, we generate t-SNE plots \citep{tsne} of the predicted embeddings of input speech in each source accent and output speech in each target accent, as depicted in \autoref{fig:tsne_plots}. We find the accent classifiers effectively cluster input speech according to their source accents and further observe that output speech generally preserves these clusters after accent conversion regardless of the source accent.

We evaluate the speaker similarity for each conversion type of our BN2BN method alongside additional reference values using the ASV metric. We find that the timbre of the original speaker is largely preserved in both monolingual and cross-lingual scenarios after accent conversion, as summarized in \autoref{ac_asv}.

\subsection{Zero-Shot Identity-Preserving Speech Style Conversion} 

We evaluate VoiceShop's speech style conversion capabilities using three styles (``Sarcastic Youth,'' ``Formal British,'' and ``Cartoon Character''). In these examples, the target styles are acquired from in-house datasets, whereby actors perform highly stylized speech, amounting to a total of approximately one hour of data per style. Conversion occurs in a many-to-one manner, such that English speakers of arbitrary source accents are converted to the target style while preserving their timbre in a zero-shot manner. We use our \emph{VS-EN-L10} backbone model to generate samples. %

\paragraph{Subjective MOS Evaluation.} To measure the perceived conversion strength of the aforementioned speech styles, we conduct a MOS study consisting of 20 anonymous native English speakers balanced across gender using the Prolific crowd-sourcing platform. We use the Microsoft Azure TTS service to generate speech in 7 timbres and convert each to the target speaking styles using our BN2BN models. We ask participants to rate the perceived similarity of a sample's speaking style against a provided reference clip on a scale from 1 to 5, where 1 indicates low similarity and 5 indicates high similarity, for three categories: ground truth samples extracted from the training corpus, neutral speech inputs, and model outputs with converted speaking styles (21 samples each). We summarize results in \autoref{sc_mos} and find the outputs of our style conversion models perform comparably well to ground truth samples, as opposed to the neutral inputs, indicating that the target speaking styles are reliably transferred to unseen speakers.

\begin{table}[tb]
\parbox{.48\linewidth}{
\caption{Results of subjective MOS evaluation of style conversion measuring \emph{conversion strength} of three styles (``Sarcastic Youth,'' ``Formal British,'' and ``Cartoon Character'').}
\centering
\label{sc_mos}
\begin{small}
\begin{tabular}{lcc}
\toprule
\textbf{Speech Type} & \textbf{Style MOS ($\uparrow$)} \\
\midrule
Non-Stylized Input & 1.57$\pm$0.10 \\
Model Output & \textbf{3.83$\pm$0.13} \\
\midrule
Same Speaker and Style & 4.05$\pm$0.12 \\
\bottomrule
\end{tabular}
\end{small}
}
\hfill
\parbox{.48\linewidth}{
\caption{Results of style conversion objective evaluation, measuring \emph{speaker similarity}. We denote the number of individual samples used to calculate each value in parentheses.}
\centering
\label{sc_asv}
\begin{small}
\setlength{\tabcolsep}{5pt}
\begin{tabular}{lc}
\toprule
\textbf{Speech Type} & \textbf{ASV ($\uparrow$)} \\
\midrule
Non-Matching Speaker & 0.168$\pm$0.012 (420) \\
Model Output & \textbf{0.492$\pm$0.015} (210) \\
\midrule
Same Speaker and Style & 0.903$\pm$0.003 (210) \\
\bottomrule
\end{tabular}
\end{small}
}
\end{table}

\paragraph{Objective Evaluation.} We again use ASV to measure the speaker similarity of our style conversion models under the same configuration and provide results in \autoref{sc_asv}. 
We note that a fair comparison should be made by calculating the ASV between the same speaker's stylized and non-stylized speech. 
However, such paired samples do not exist as the ground truth. 
Therefore, the ASV score for \emph{same speaker and style} in \autoref{sc_asv} does not expose the ASV drop due to the stylization of the same person's speech, which is reflected in the score for our \emph{model output}. 

\subsection{Zero-Shot Combined Multi-Attribute Editing} 
Rather than performing serial editing of individual attributes, we showcase our framework's performance on combined multi-attribute editing, whereby arbitrary unseen users can modify their accent, age, and gender simultaneously in a single forward synthesis pass while preserving their timbre. This capability is enabled by our plug-in modular design that allows concurrent editing on both the speaker embedding and content features of our \emph{VS-EN-CN-L18} backbone model.

\paragraph{Subjective CMOS Evaluation.} We design a modified CMOS study to measure two key qualities of VoiceShop's multi-attribute editing capabilities: overall editing capabilities, i.e., participants agree that attributes we claim to edit are in fact edited according to their best judgment, and attribute disentanglement, i.e., the modification of any combination of attributes does not alter participants' perception of other attributes we claim are kept constant. 

To this end, we generate samples using two source speakers and perform all permutations of multi-attribute editing for one, two, and three attributes (i.e., edit accent, age, and gender individually, edit accent and age, accent and gender, and age and gender simultaneously, and edit all three attributes simultaneously). For each transformation, participants are provided with input and output speech pairs and are asked to identify which of the samples is more accurately described by a text description for each attribute, where they may select ``neither'' if they believe neither sample is described by the text (e.g., if given a sample of a male speaker where we only perform accent conversion, we would like participants to correctly select ``neither'' if asked which sample sounds like it was more likely spoken by a female speaker). An additional advantage of including ``neither'' as a response is that listeners are not obliged to give answers favorable to our system if the editing strength is not strong enough to warrant a clear preference. This design ensures that participants not only agree with the correct descriptions of provided samples, but also that they can confidently identify when the descriptions are incorrect.

We recruit 20 anonymous native English speakers balanced across genders using the Prolific crowd-sourcing platform and summarize our findings in \autoref{multi_attribute_results} of 42 input-output pairs. We observe the majority of participants can correctly identify when any attribute has or has not been edited, suggesting that our approach is capable of strong, disentangled combined multi-attribute editing in a  single forward pass. 

\begin{table}[t]
\caption{Results of multi-attribute editing subjective CMOS evaluation. The accuracy indicates the proportion of participants who correctly completed each task. We denote the number of responses for each task in parentheses. Some tasks have overlapping samples.}
\label{multi_attribute_results}
\begin{center}
\begin{small}
\setlength{\tabcolsep}{5pt}
\begin{tabular}{lp{6.5cm}cc}
\toprule
\textbf{Category} & \textbf{Task} & \textbf{Accuracy ($\uparrow$)} \\
\midrule
\multirow{6}{*}{\shortstack[l]{Correctly observe\\when an attribute\\\textbf{has} been edited}} & Correctly observe \textbf{accent} editing & 89.38\% (160) \\
\cmidrule{2-4}
& Correctly observe \textbf{age} editing & 92.50\% (160) \\
\cmidrule{2-4}
& Correctly observe \textbf{gender} editing & 96.88\% (160) \\
\cmidrule{2-4}
& Correctly observe that \textbf{any attribute} has changed, regardless of which attributes are edited & \multirow{2}{*}{92.92\% (480)} \\
\midrule
\multirow{9}{*}{\shortstack[l]{Correctly observe\\when an attribute\\\textbf{has} \textbf{not} been edited}} & Correctly observe other attributes have \textbf{not changed} when editing \textbf{accent} & \multirow{2}{*}{80.62\% (160)} \\
\cmidrule{2-4}
& Correctly observe other attributes have \textbf{not changed} when editing \textbf{age} & \multirow{2}{*}{99.38\% (160)} \\
\cmidrule{2-4}
& Correctly observe other attributes have \textbf{not changed} when editing \textbf{gender} & \multirow{2}{*}{86.25\% (160)} \\
\cmidrule{2-4}
& Correctly observe other attributes have \textbf{not changed} when editing \textbf{any attribute} & \multirow{2}{*}{87.78\% (360)}\\
\midrule
\multirow{2}{*}{Overall accuracy} & Correctly observe when \textbf{any attribute} \textbf{has} or \textbf{has not} been edited & \multirow{2}{*}{90.71\% (840)} \\
\bottomrule
\end{tabular}
\end{small}
\end{center}
\end{table}

\begin{figure}[b]
\centering
\begin{subfigure}{0.49\linewidth}
\includegraphics[width=\linewidth]{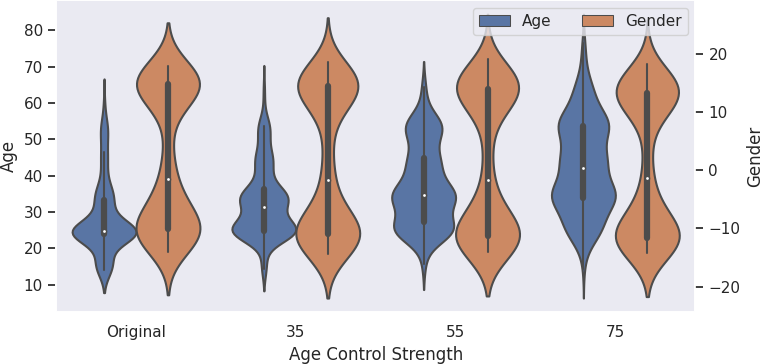}
\caption{Age editing}
\end{subfigure}
\begin{subfigure}{0.49\linewidth}
\includegraphics[width=\linewidth]{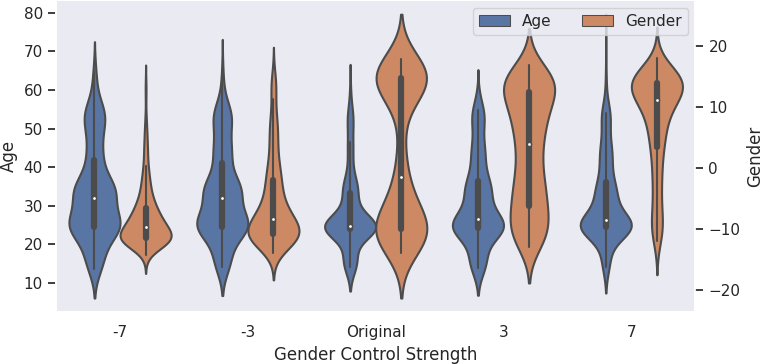}
\caption{Gender editing}
\end{subfigure}
\caption{{\bf Predicted age and gender distributions after}: (a) editing age only; (b) editing gender only. We show that not only does the edited attribute shift to the desirable direction with varying editing signal strength but also the unedited attribute remains similar.}
\label{fig:age-gender-editing}
\end{figure}

\paragraph{Objective Evaluation.} We first experiment by editing age and gender using out-of-domain speech samples from VCTK by randomly selecting three speech samples for each of the 109 speakers in this dataset, which are also used by NANSY++ \citep{choi2022nansy++}. We use the attribute predictor in \S\ref{sec:attribute-cnf} to analyze the age and gender distributions of VCTK samples before and after our attribute editing, with results shown in~\autoref{fig:age-gender-editing}. We demonstrate that our framework achieves competitive age and gender editing strength, despite that our CNF is only trained on weakly-supervised public attribute datasets. For this comparison, we use our \emph{VS-EN-L18} backbone model.

\section{Conclusion}

In this work, we presented VoiceShop, a novel speech-to-speech framework that enables the modification of multiple speech attributes while preserving the input speaker's timbre in both monolingual and cross-lingual settings. It overcomes the limitations of previous models by enabling simultaneous editing of attributes, providing zero-shot capability for out-of-domain speakers, and avoiding timbre leakage that alters the speaker's unedited attributes in voice editing tasks. It additionally provides a new assortment of methods to tackle the under-explored voice editing task.

While VoiceShop offers new capabilities for fine-grained speaker attribute editing, an apparent limitation of our work is that downstream tasks are still bounded by the quality of supervised data. For example, due to the highly imbalanced age distribution of our CNF attribute dataset, editing performance becomes limited as we reach under-represented age ranges, and the perceived naturalness of our BN2BN output is capped by that of the synthetic speech used for training. Furthermore, while we showcase cross-lingual synthesis capabilities, we are still constrained to English and Mandarin content, motivating the exploration of universal speech representations that generalize to all languages, which we leave to future work.

\section{Ethical Considerations}

As with all generative artificial intelligence systems, the real-world impact and potential for unintended misuse of models like VoiceShop must be considered. While there are beneficial use cases of our framework, such as providing entertainment value or lowering cross-cultural communication barriers by allowing users to speak other languages or accents in their own voice, its zero-shot capabilities could enable a user to generate misleading content with relative ease, such as synthesizing speech in the voice of an individual without their knowledge, presenting a risk of misinformation. This concern is not unique to VoiceShop and motivates continued efforts towards developing robust audio deepfake detection models deployed in tandem with generative speech models \citep{guo2024audio, zang2024singfake, zhang2023remember} or the incorporation of imperceptible watermarks in synthesized audio signals to classify genuine or generated speech \citep{cao2023invisible, chen2024wavmark, juvela2024collaborative, liu2023detecting, roman2024proactive}, although this subject is outside the scope of this work. In an effort to balance the need for transparent, reproducible, and socially responsible research practices, and due to the proprietary nature of portions of data used in this work, we share the details of our findings here, but do not plan to publicly release the model checkpoints or implementation at this time. The authors do not condone the use of this technology for illegal or malicious purposes.

\bibliographystyle{plainnat}
\bibliography{refs}

\newpage
\appendix
\counterwithin{figure}{section}
\counterwithin{table}{section}
\renewcommand\thefigure{\thesection\arabic{figure}}
\renewcommand\thetable{\thesection\arabic{table}}
\onecolumn
\section{Appendix}

\subsection{Implementation details of conformer-based ASR models}
\label{sec:asr_details}

\begin{figure}[t]
\centering
\includegraphics[width=0.9\linewidth]{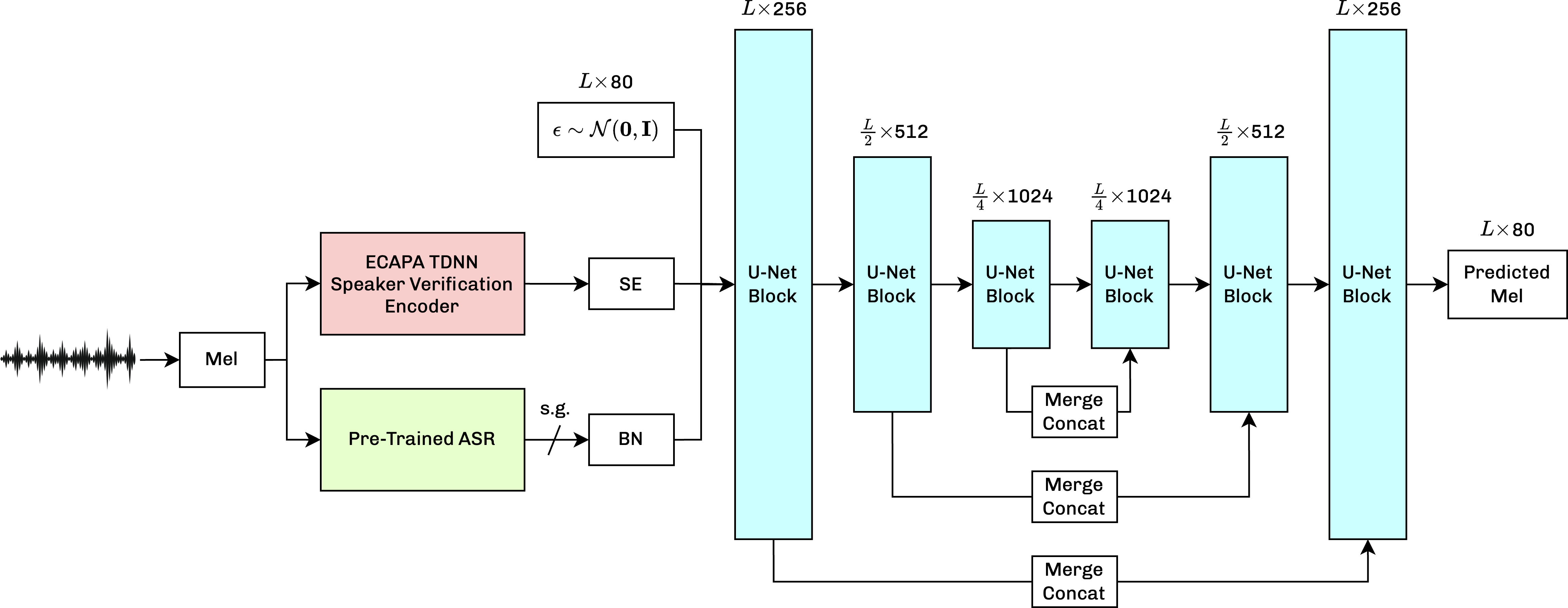}
\caption{The architecture of our conditional diffusion backbone model, consisting of an ECAPA-TDNN-based speaker encoder \citep{desplanques2020ecapa} and one-dimensional U-Net \citep{ronneberger2015unet} inspired by Moûsai \citep{schneider2023mousai}.}
\label{diff-u-net}
\end{figure}

We use the open-source ESPnet\footnote{\url{https://github.com/espnet/espnet}} \citep{watanabe2018espnet} library to train our ASR model, following the ESPnet2 recipe with a conformer-based encoder and a transformer-based decoder. We add an additional head to the output of the conformer encoder to calculate logits for a Connectionist Temporal Classification (CTC) \citep{10.1145/1143844.1143891} loss. The encoder and decoder are connected with cross attention, as in the standard transformer network design \citep{NIPS2017_3f5ee243}. We optimize the model by jointly minimizing CTC and cross-entropy (CE) loss terms on the output of the transformer decoder, using a hyperparameter $\lambda$ to interpolate the weights of the two terms. Given paired ASR training data of speech and corresponding text $(X, Y)$, we define the training objective as follows:
\begin{equation}
\label{asr_training}
\mathcal{L} = \lambda \log p_{\text{CTC}}(Y|X) + (1-\lambda)\log p_{\text{Dec}}(Y|X)
\end{equation}
Since their debut, conformers \citep{Gulati_2020} have been widely adopted in the ASR community. Compared to conventional transformer encoders, conformers add a convolutional module between the multi-head self-attention and feed-forward module in each computational block, providing the ability to model the local interactions between speech representations besides the global modeling ability enabled by multi-head self-attention. Thus, to learn better semantic representations for VoiceShop, we adopt a conformer-based encoder for our ASR model.

We train monolingual and bilingual versions of the ASR model, referred to as \emph{ASR-EN} and \emph{ASR-EN-CN} respectively, such that the former only transcribes English speech and the latter transcribes both English and Mandarin speech. The conformer encoder of both configurations consists of 16 blocks and the transformer decoder respectively contains 6 and 4 blocks in \emph{ASR-EN} and \emph{ASR-EN-CN}. \emph{ASR-EN} uses a hidden dimension of 768 for both encoder and decoder modules, containing approximately 220M learnable parameters, whereas \emph{ASR-EN-CN} sets this value to 512 with roughly 140M total parameters. Both models set $\lambda$ in \autoref{asr_training} to 0.1 and apply SpecAugment \citep{Park_2019} to improve their generalization ability.

\emph{ASR-EN} is trained on an in-house dataset consisting of 40K hours of English pairs of audio and corresponding transcriptions provided by human annotators, amounting to 4M utterances, while \emph{ASR-EN-CN} is trained on an in-house dataset consisting of 60K hours of in-house Mandarin data and 5K hours of in-house English data, totaling 5M utterances. We use a joint vocabulary of 11K unique tokens consisting of English subwords prepared using the SentencePiece tokenizer which implements Byte-Pair Encoding (BPE) and Mandarin characters, and omits Mandarin tokens for the monolingual model. Both models are trained using the Adam optimizer \citep{kingma2014adam} on 32 V100 GPUs with a learning rate $5 \times 10^{-4}$, 20K warm up steps, and a batch size equivalent to 2 hours of audio data for 40 epochs. \emph{ASR-EN} achieves a word error rate (WER) of 8.86\% on an internal English test set and \emph{ASR-EN-CN} achieves a character error rate (CER) of 7.4\% on an internal Chinese test set in the video domain.

\subsection{Implementation details of conditional diffusion backbone model}
\label{sec:diffusion_details}

We modify the text-to-music framework Moûsai \citep{schneider2023mousai} for the design of our conditional diffusion backbone model, depicted in \autoref{diff-u-net}, and describe its implementation details in this section. We base our code on their open-source implementation\footnote{\url{https://github.com/archinetai/audio-diffusion-pytorch}} to enable better reproducibility.

Given a mel-spectrogram of speech, the diffusion model accepts two conditioning signals: an utterance-level global speaker embedding produced by a speaker encoder and a time-varying content embedding produced by a pre-trained ASR or APR model. The speaker encoder is trained jointly with the diffusion model from scratch and adopts the same configuration as the ECAPA-TDNN speaker verification model \citep{desplanques2020ecapa}, whereas the weights of the ASR or APR model remain frozen during training.

\begin{figure}[h] %
\centering
\includegraphics[width=0.6\linewidth]{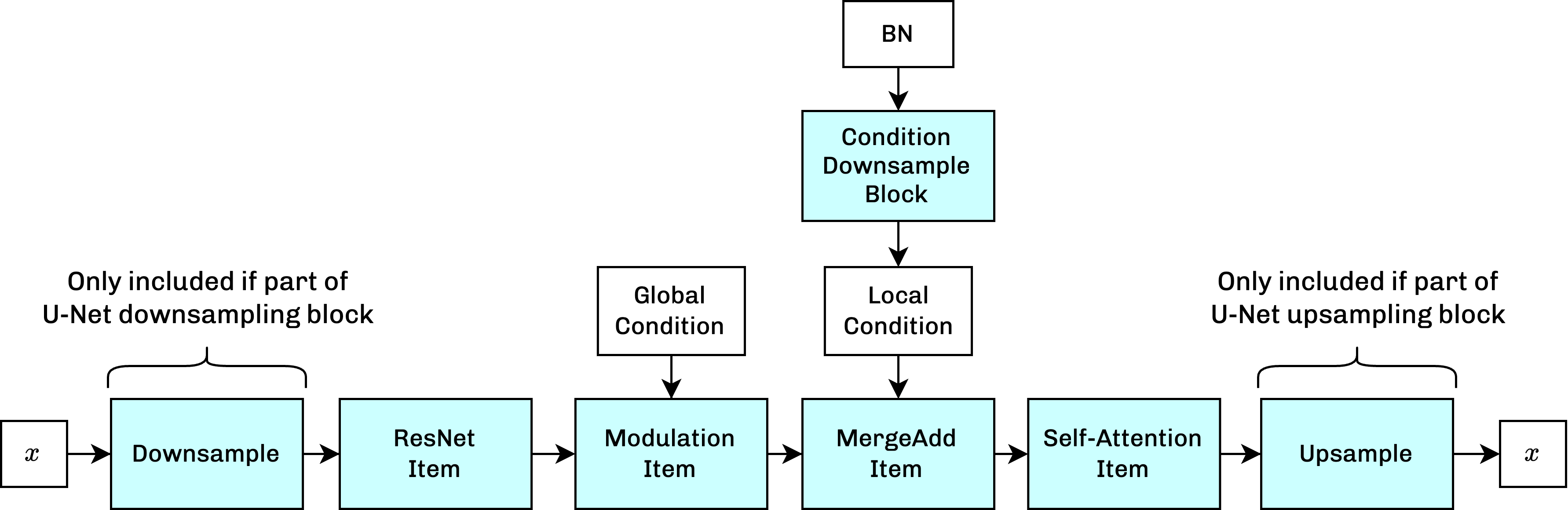}
\caption{The design of the U-Net blocks of our diffusion backbone model.}
\label{diff-u-net-block}
\end{figure}

\begin{figure*}[h]
\centering
\includegraphics[width=\linewidth]{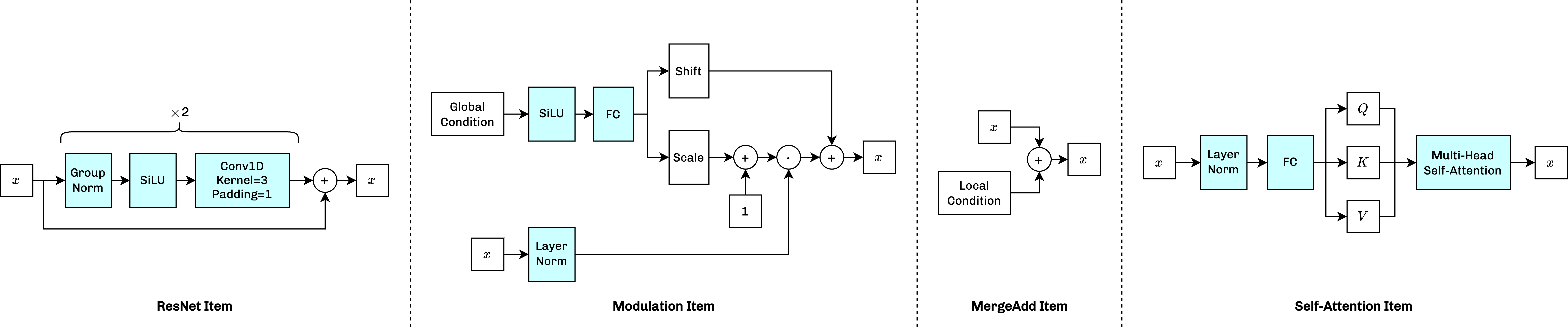}
\caption{Details of each ``U-Net item'' used to process noise-injected samples and incorporate conditioning signals into the model.}
\label{diff-items}
\end{figure*}

We adopt the one-dimensional U-Net \citep{ronneberger2015unet} of \citet{schneider2023mousai} as the architecture of our diffusion model and follow their terminology of ``U-Net items'' to describe the various components in each U-Net block which incorporate conditioning signals into the model. Please refer to \autoref{diff-u-net-block} for a visual depiction of our modified U-Net block design. Each U-Net block sequentially processes data in a series of operations consisting of a one-dimensional convolutional ``ResNet item''  \citep{he2015deep}, a ``modulation item'' where the time step and global speaker embeddings are provided to the model, a ``MergeAdd item'' where the time-varying content features are supplied, and a ``self-attention item.'' Note that we remove the ``cross-attention item'' included in their work, as our framework does not use text, and replace their ``inject item'' with a simpler piecewise addition operation. The details of each item are depicted in \autoref{diff-items}.

\begin{figure*}[t]
\centering
\includegraphics[width=\linewidth]{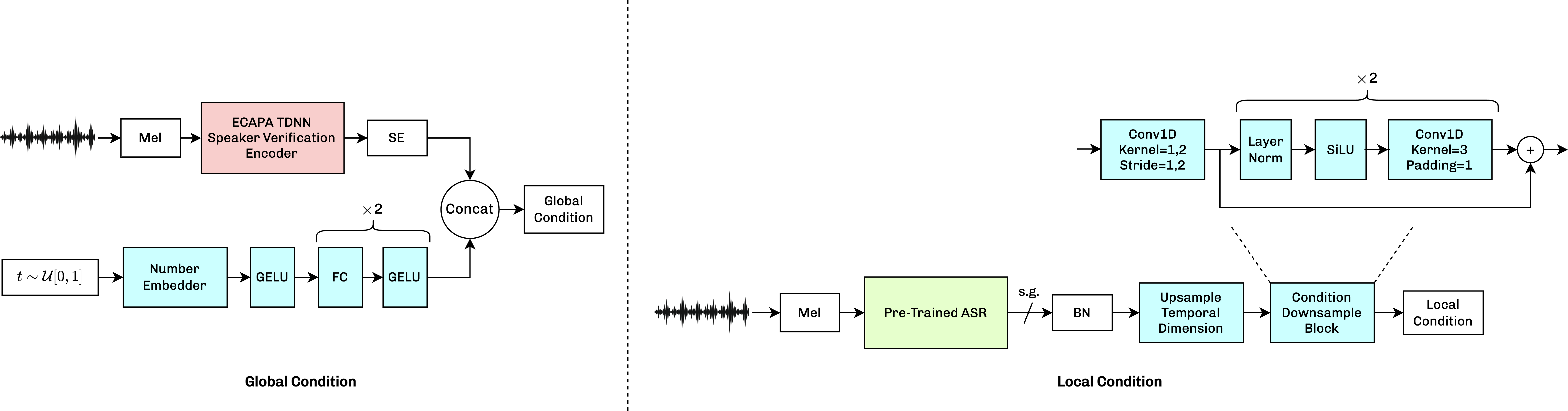}
\caption{Details of the preparation of ``global'' and ``local'' conditions, the first of which consists of the concatenation of the utterance-level speaker embedding and time step embedding and the second of which consists of the upsampled time-varying content features following additional convolutional processing with residual connections.}
\label{diff-conditions}
\end{figure*}

As shown in \autoref{diff-conditions}, the speaker embedding is concatenated with the time step embedding of the current iteration $t$ of the reverse generative process, while the time-varying content features are upsampled along the temporal dimension to match that of the mel-spectrogram and undergo additional convolutional layers with residual connections. We respectively refer to the outputs of these operations as the ``global'' and ``local'' conditions. 

\begin{figure*}[h]
\centering
\includegraphics[width=0.3\linewidth]{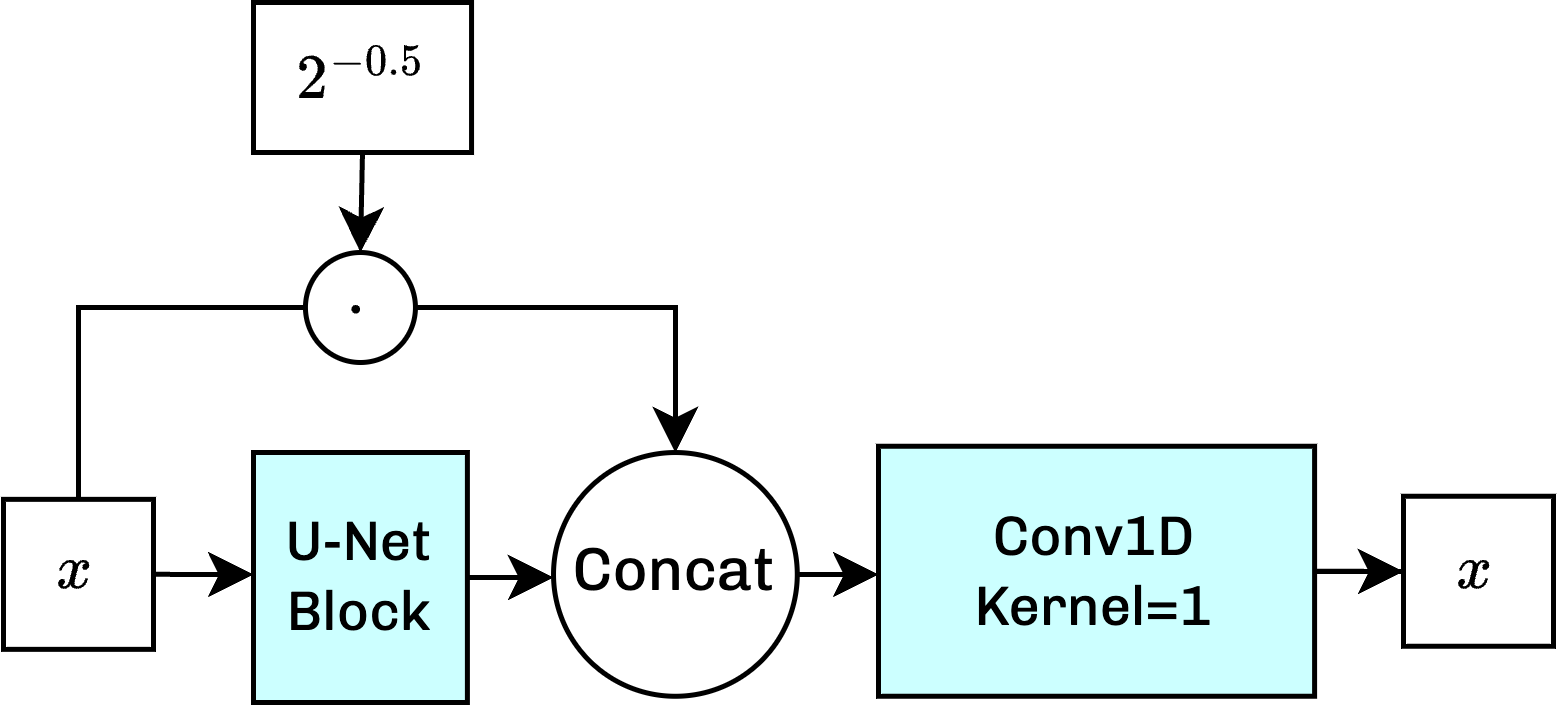}
\caption{Details of the ``MergeConcat'' operation.}
\label{diff-merge_concat}
\end{figure*}

A noise-injected sample is first processed by the ``ResNet item,'' after which the global condition is provided to the ``modulation item,'' where it is used to predict the parameters of an affine transformation applied to the output of the ``ResNet item.'' The local condition is then passed to the ``MergeAdd item,'' where it is added directly to the output of the ``modulation item.'' Finally, the sample is passed through a linear projection layer, which predicts query, key, and value tensors received by a standard multi-head self-attention layer \citep{NIPS2017_3f5ee243}. The residual connections of the U-Net are applied using a ``MergeConcat'' operation depicted in \autoref{diff-merge_concat}.

\subsection{Implementation details of vocoder}
\label{sec:vocoder_details}
We adopt HiFi-GAN \citep{kong2020hifigan} as our vocoder for its low latency and competitive performance against autoregressive alternatives such as WaveNet \citep{oord2016wavenet}, but with key modifications to its architecture and objective function. To address occasional, undesirable glitches in speech synthesized by pre-trained instances of HiFi-GAN, we add a learnable sinusoidal activation module inspired by the source-filter theory of speech production \citep{fant1971acoustic}, which receives mel-spectrograms of speech and predicts its corresponding excitation signal representing the fundamental frequency ($F_0$) contour of an utterance to be provided as extra conditioning signals to the convolutional blocks of the generator and discriminator modules. We make use of the originally proposed multi-period and multi-scale discriminators, adding integrated sub-band coding filters to reduce aliasing, and reduce the receptive field of HiFi-GAN's convolutional layers to correspond to 7 frames for spectrograms with a hop size of 240.

We employ HiFi-GAN's mel-spectrogram regression and feature-matching loss terms, as follows: 
\begin{align}
\mathcal{L}_\text{Mel} &=\mathbb{E}_{(x,s)}\big[ \| \phi(x) - \phi(G(s)) \| \big] \\
\mathcal{L}_\text{FM}(G;D)&=\mathbb{E}_{(x_s)} \left[ \sum_{i=1}^T \frac{1}{N} \| D^i(x) -D^i(G(s)) \| \right]
\end{align}
such that $\phi$ is the function that converts time-domain waveforms to their mel-spectrogram representations. Furthermore, we add the energy, time, and phase loss terms proposed by TFGAN \citep{tian2020tfgan}, which we find assists the model converge relatively earlier than when omitted, defined as:
\begin{align}
\mathcal{L}_\text{Energy} &= \| \mathbb{E}(x^2) - \mathbb{E}(\hat{x}^2) \|_1 \\
\mathcal{L}_\text{Time} &= \| \mathbb{E}(x) - \mathbb{E}(\hat{x}) \|_1 \\
\mathcal{L}_\text{Phase} &= \| \Delta x - \Delta \hat{x} \|_1
\end{align}
such that $x$ and $\hat{x}$ correspond respectively to the ground truth and synthesized audio signals, $\Delta$ refers to the first-difference estimator, and $\|\cdot\|_1$ is the $L^1$ Frobenius norm. We train the sinusoidal excitation module using MAE on the ground truth and predicted $F_0$ contours, $s$ and $\hat{s}$, respectively:
\begin{equation}
\mathcal{L}_{F_0} = \mathbb{E}_s \big[ \| \ln{s} - \ln{\hat{s}} \| \big]
\end{equation}
We remove the originally proposed LSGAN loss, which we empirically find improves the stability of the adversarial training procedure. The final loss function for our modified HiFi-GAN vocoder is formulated as:
\begin{equation}
\begin{split}
\mathcal{L}_\text{Vocoder} = \lambda_1 \mathcal{L}_\text{Mel} + \lambda_2 \mathcal{L}_\text{FM} + \lambda_3 \mathcal{L}_\text{Energy} + \lambda_4 \mathcal{L}_\text{Time} + \lambda_5 \mathcal{L}_\text{Phase} + \lambda_6 \mathcal{L}_{F_0}
\end{split}
\end{equation}
such that $\lambda_1=1, \lambda_2=1, \lambda_3=100, \lambda_4=200, \lambda_5=100$, and $\lambda_6=1$. We train the vocoder using the Adam optimizer \citep{kingma2014adam} for the generator and all discriminators, each with a learning rate of $1 \times 10^{-3}$, providing the generator with a warm up of 2K iterations. The model is trained for 2M iterations on 8 V100 GPUs using a batch size of 12 samples on roughly 409 hours of in-house multilingual speech data consisting primarily of English and Mandarin speech, and to a lesser extent additional languages, such as Japanese and Indonesian, after which it robustly generates 24kHz audio. 

\begin{figure*}[t] 
\centering
\includegraphics[width=\linewidth]{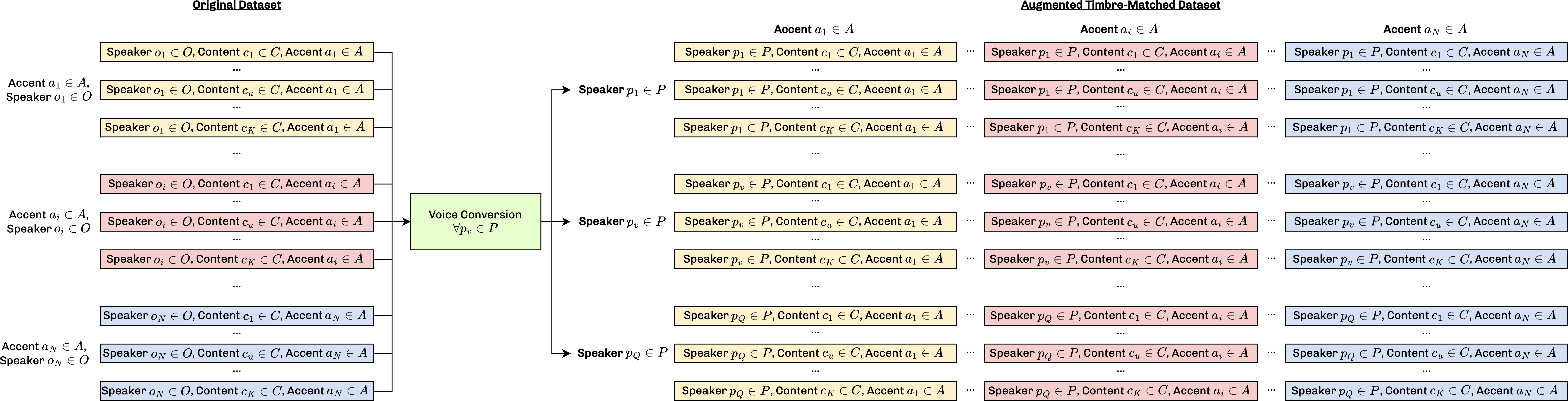}
\caption{{\bf BN2BN dataset preparation:} Given accents $ A = \{ a_i \}_{i=1}^N$, content  $C =\{ c_u \}_{u=1}^K$, original speakers $ O = \{ o_i \}_{i=1}^N$, and target speakers $ P =  \{ p_v \}_{v=1}^Q$, we propose a simple and intuitive method of transforming widely available non-parallel datasets into parallel datasets for accent and speech style conversion through TTS and VC augmentation.}
\label{parallel}
\end{figure*}

\subsection{Augmenting non-parallel speech into parallel ``timbre-matched'' datasets}
\label{sec:timbre-matched}

Our proposed bottleneck-to-bottleneck (BN2BN) modeling approach necessitates the use of a parallel training dataset consisting of audio samples of individuals speaking the same content in a variety of accents or speech styles. The collection of recorded speech that meets such requirements is typically understood to be preventatively difficult, time-consuming, and costly in existing literature. One apparent drawback of manually recording such a dataset, apart from other practical concerns, is that it requires a single person to fluently speak in a variety of accents, most of which will not be native to the speaker, which can result in inaccurate depictions of the accents when instead it would be preferable to collect the speech of native speakers of the various accents to achieve faithful accent and speech style conversion. By leveraging the rapid progress of neural TTS and VC modeling available in a wide array of accents, we demonstrate that acquiring such a dataset can be straightforward, efficient, and cost-effective. In fact, while it is not necessary, our proposed approach can be trained entirely on synthetically generated speech. 

As our BN2BN approach belongs to the family of attention-based sequence-to-sequence models, we do not require that speech samples in different accents or speech styles have the same duration, i.e., inputs and outputs may differ in length. Furthermore, we do not require explicit frame-level alignments. In practice, we have empirically observed that high-quality many-to-many accent or speech style conversion can be achieved using our BN2BN model with as little as one hour of speech for a given accent or style. For unusual speech styles, given a textual transcription of a speaker's recordings, one may use TTS models to generate accompanying parallel data of the same content in a variety of accents, after which we propose the use of VC models to align the timbres of the speech samples across all accents or styles, culminating in a multi-speaker, multi-accent dataset which may be used to train BN2BN models. We describe the process of preparing a parallel dataset for this task in detail below. For simplicity, we will only refer to accent conversion for the following explanation, but we apply the same set of procedures to achieve generalized speech style conversion as well.

Assume there exists a set $A$ of $N$ accents to be modeled, such that $A=\{a_i\}_{i=1}^N$. We assume that each accent $a_i$ corresponds to its own original native speaker $o_i$, such that the set of original speakers may be defined as $O=\{o_i\}_{i=1}^N$ and $|A|=|O|$, although in practice this may not necessarily be the case. For all accents $a_i\in A$,  we obtain audio samples of the original speakers $o_i\in O$ speaking from the same shared text corpus $C$ consisting of $K$ total utterances, such that $C=\{c_u\}_{u=1}^K$ (e.g., the content of $c_k$ may correspond to the sentence, ``Today's weather forecast promises intermittent showers with periods of sunshine,'' which will be spoken in all $N$ accents in $A$). The source of these audio samples may be real recordings of humans or synthetically generated using TTS models. Note that while we now possess audio samples of the same content being spoken in various accents, each accent corresponds to different timbres. Therefore, we perform inference with pre-trained VC models to convert the timbres of the original speakers $o_i\in O$ across all accents to a new set $P$ of $Q$ target speakers, such that $P=\{t_v\}_{v=1}^Q$. After doing so, we have completed preparing a parallel dataset that satisfies our initial requirements for this task: for each target speaker $p_v\in P$, we possess audio samples of the same corpus of utterances $C$ being spoken in all accents $a_i\in A$. Refer to \autoref{parallel} for a visual depiction of this parallel dataset preparation, the result of which we refer to as the ``timbre-matched dataset.''

\begin{figure*}[t] 
\centering
\includegraphics[width=\linewidth]{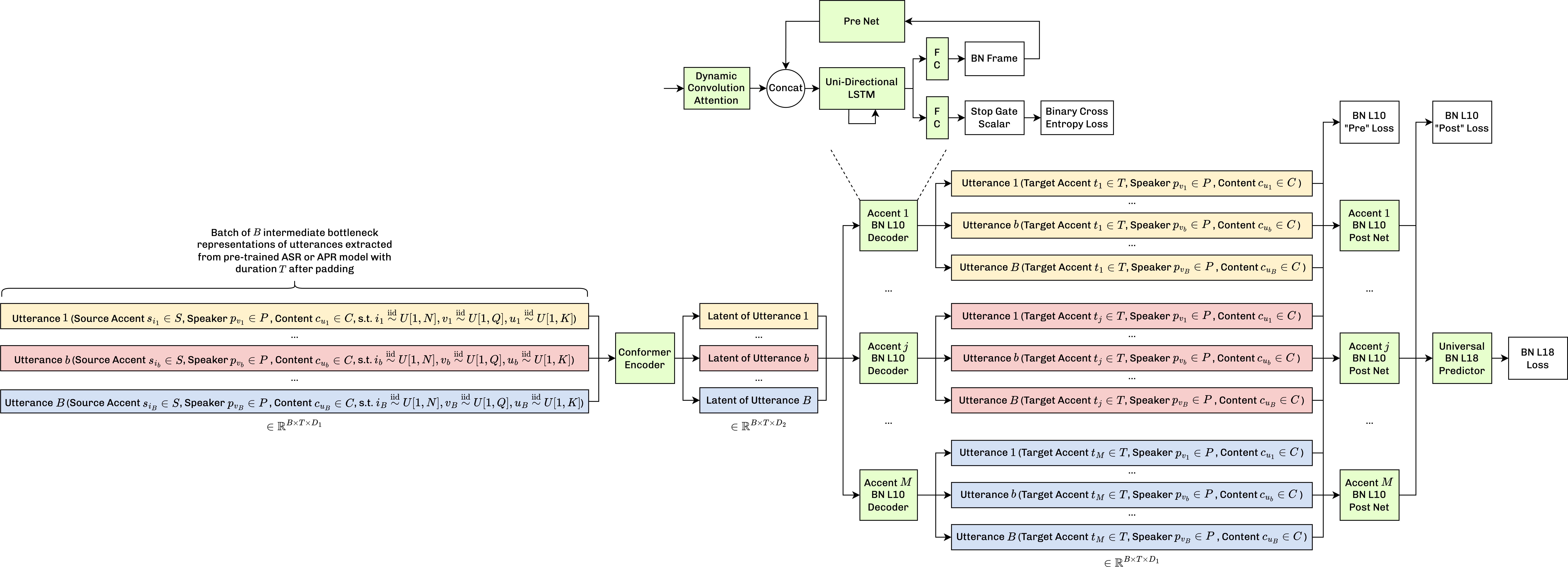}
\caption{{\bf Details of bottleneck-to-bottleneck (BN2BN) modeling:} Given source accents $ S = \{ s_i \}_{i=1}^N $, target accents $ T = \{ t_j \}_{j=1}^M$, speakers $ P =\{ p_v \}_{v=1}^Q$, and content $ C =\{ c_u \}_{u=1}^K$, our BN2BN design maps the time-varying content features of utterances from an arbitrary number of source accents to those of an arbitrary number of target accents in a single model using a multi-decoder architecture.}
\label{bn2bn_details}
\end{figure*}

For each audio sample of the timbre-matched dataset, we first apply the Short-Time Fourier Transform (STFT) to obtain a time-frequency representation of the raw time-domain speech signal in the form of a mel-spectrogram. We then use pre-trained ASR or APR models to extract time-varying embeddings that represent the content of the speech signal. Note that our approach does not impose that a specific type of content feature be extracted, as this method is generalizable to any time-varying content representation that can be assumed to encode accent or speech style information in its learned latent embedding space. Additional training details of the BN2BN model are depicted in \autoref{bn2bn_details}.

For this work, we train three BN2BN models, a monolingual English model, a monolingual Mandarin model, and a cross-lingual model, the last of which combines both languages. To prepare our ``timbre-matched'' parallel dataset, we convert each accent's original synthetic speech to 6 target timbres for English models and 7 target timbres for Mandarin models, where each speaker's accent subset contains roughly 4.7 hours of data. Our monolingual English model is trained in a many-to-many manner consisting of the following accents: American, Australian, British, Chinese, French, Greek, Indian, Italian, Kenyan, Korean, Mexican, and Russian. Our monolingual Mandarin model is trained in an identical manner on the following accents: Guangxi, Henan, Jilu, Liaoning, Mandarin, Shaanxi, Sichuan, and Taiwanese. Our cross-lingual model combines languages and contains American, British, French, Indian, Kenyan, Jilu, Mandarin, Shaanxi, Sichuan, and Taiwanese accents. Note that in the cross-lingual case, we do not have parallel data of English content in Chinese accents or Mandarin content in English accents, but we find we can still successfully transfer these accents across languages while faithfully preserving the source content using our BN2BN framework. The specific public TTS models used to generate the ground truth accent data can be found in \autoref{tts_models}.

\subsection{Differences to ``BN2BN'' module proposed by \citet{accentspeech}}

In a separate and independent work, \citet{accentspeech} also proposes the use of a learnable module referred to as ``BN2BN'' to achieve accent transfer within a TTS framework named AccentSpeech. While both works share a similar intuition of training a BN2BN model to map between the bottleneck features of utterances in a source accent and those of utterances in a target accent, we distinguish our approach from theirs by noting the following key differences in our proposed methods. 

AccentSpeech is a TTS model, whereas VoiceShop is STS, only requiring audio as input to convert the accent or speech style of a given utterance. Unlike VoiceShop, the primary problem AccentSpeech addresses is how to leverage lower quality crowd-sourced data to enable cross-speaker accent transfer to a single ``unaccented'' target speaker, whereas our model aims to achieve identity-preserving many-to-many accent and speech style conversion in a single framework. AccentSpeech is capable of one-to-one accent transfer, and therefore requires training separate BN2BN models for each desired conversion path, e.g., Mandarin-to-Chengdu and Mandarin-to-Xian must be trained as two individual models. VoiceShop is capable of many-to-many accent and speech style transfer, allowing a single BN2BN model to learn the mappings between all conversion paths across source accents and speech styles simultaneously.

Since the decoder is trained on the high-fidelity recordings of a single speaker, AccentSpeech is capable of synthesizing accented speech in one target timbre, whereas VoiceShop aims to preserve any arbitrary source speaker's timbre in the conversion process. AccentSpeech uses feed-forward transformer blocks stacked between one-dimensional convolutional layers as the model design of its BN2BN module, while we adopt an attention-based sequence-to-sequence architecture. One disadvantage of this design choice on the part of \citet{accentspeech} is that AccentSpeech needs frame-wise time-aligned parallel data between two speakers of different accents, requiring third-party forced alignment tools. Another consequence of this restriction is that duration must be considered irrelevant to accent information. VoiceShop does not require frame-level alignments for parallel speech pairs and allows such pairs to be of different durations. Furthermore, VoiceShop is capable of modeling the varying rhythmic patterns of native speech across accents and speech styles, which, in our view, is an important factor to the perceived fluency of our model's output speech.

\subsection{Training configurations, model hyperparameters, and additional evaluation results}
\label{sec:appendix_additional}

We use the following section to provide the precise values of hyperparameters for various models to enable better reproducibility of our results, such as the BN2BN accent and speech style conversion model, conditional diffusion backbone model, and mel-spectrogram extraction settings. We also provide additional evaluation results for various synthesis tasks, as discussed in the main body.

\begin{table*}[b]
\caption{BN2BN model hyperparameters, where values corresponding to BN2BN models trained on content features extracted from \emph{ASR-EN} are colored in {\color{red}red} and values corresponding to those of \emph{ASR-EN-CN} are colored in {\color{blue}blue}. The optional adversarial domain adaptation classifier is only included in cross-lingual accent conversion settings when explicitly specified.}
\label{bn2bn_hparam}
\begin{center}
\begin{small}
\setlength{\tabcolsep}{5pt}
\begin{tabular}{cll}
\toprule
\textbf{Module} & \textbf{Hyperparameter} & \textbf{Value} \\
\midrule
\multirow{10}{*}{\shortstack{Universal\\Conformer\\Encoder}} & Encoder Dimension & 512 \\
 & Layers & 12 \\
 & Attention Heads & 8 \\
 & Feedforward Expansion Factor & 3 \\
 & Convolutional Expansion Factor & 2 \\
 & Input Dropout Probability & 0.0 \\
 & Feedforward Dropout Probability & 0.1 \\
 & Convolutional Dropout Probability & 0.1 \\
 & Convolutional Kernal Size & 9 \\
 & Half Step Residual & True \\
 \hline
\multirow{9}{*}{\shortstack{BN L10\\Tacotron 2\\Decoder}} & Decoder Dimension & {\color{red}768}, {\color{blue}512} \\
 & Frames Per Step & 1 \\
 & Attention RNN Layers & 2 \\
 & Attention RNN Dimension & 512 \\
 & Attention Zoneout Probability & 0.1 \\
 & Attention Dropout Probability & 0.1 \\
 & Stop Gate Threshold & 0.5 \\
 & Pre Net Dimension & 256 \\
 & Pre Net Dropout Probability & 0.5 \\
 \hline
\multirow{9}{*}{\shortstack{Cross-Attention\\Mechanism}} & Attention Type & Dynamic Convolution Attention\\
 & Attention Dimension & 512 \\
 & Prior Length & 10 \\
 & $\alpha$ & 0.1 \\
 & $\beta$ & 0.9 \\
 & Dynamic Channels & 8 \\
 & Dynamic Convolutional Kernal Size & 21 \\
 & Static Channels & 8 \\
 & Static Convolutional Kernal Size & 21 \\
 \hline
\multirow{4}{*}{\shortstack{BN L10\\Post Net}} & Post Net Dimension & {\color{red}768}, {\color{blue}512}\\
 & Layers & 3 \\
 & Convolutional Channels & 512 \\
 & Convolutional Kernal Size & 5 \\
 \hline
\multirow{10}{*}{\shortstack{Universal\\BN L18\\Predictor}} & Predictor Dimension & {\color{red}768}, {\color{blue}512} \\
 & Layers & 3 \\
 & Attention Heads & 8 \\
 & Feedforward Expansion Factor & 3 \\
 & Convolutional Expansion Factor & 2 \\
 & Input Dropout Probability & 0.1 \\
 & Feedforward Dropout Probability & 0.1 \\
 & Convolutional Dropout Probability & 0.1 \\
 & Convolutional Kernal Size & 9 \\
 & Half Step Residual & True \\
 \hline
 \multirow{9}{*}{\shortstack{Optional\\Adversarial\\Domain\\Adaptation\\Classifier}} & Convolutional Layers & 6 \\
 & Convolutional Channels & [32, 32, 64, 64, 128, 128] \\
 & Convolutional Kernel Size & 3 \\
 & Convolutional Stride & 2 \\
 & GRU Layers & 1 \\
 & GRU Units & 128 \\
 & Gradient Reversal $\lambda$ & -1.0 \\
 & Multilayer Perceptron Layers & 3\\
 & Multilayer Perceptron Dimension & 256\\
\bottomrule
\end{tabular}
\end{small}
\end{center}
\vskip -0.1in
\end{table*}

\begin{table*}[h]
\caption{Conditional diffusion backbone model hyperparameters, where values corresponding to diffusion models trained on content features extracted from \emph{ASR-EN} are colored in {\color{red}red} and values corresponding to those of \emph{ASR-EN-CN} are colored in {\color{blue}blue}. We only include values of downsampling blocks to avoid redundancy, such that upsampling blocks use the same values in reverse order.}
\label{diff_hparam}
\begin{center}
\begin{small}
\setlength{\tabcolsep}{5pt}
\begin{tabular}{ll}
\toprule
\textbf{Hyperparameter} & \textbf{Value} \\
\midrule
Diffusion Objective & Velocity \\
Downsampling U-Net Blocks & 3 \\
Downsampling Channels & [256, 512, 1024] \\
Downsampling Factors & [1, 2, 2] \\
Speaker Embedding Dimension & 512 \\
Time Embedding Dimension & 768 \\
Local Condition Dimension & {\color{red}768}, {\color{blue}512} \\
Attention Dimension & 512 \\
Attention Heads & 8 \\
ResNet Groups (GroupNorm) & 8 \\
Input Channels & 80 \\
Output Channels & 80 \\
\bottomrule
\end{tabular}
\end{small}
\end{center}
\vskip -0.1in
\end{table*}

\begin{table*}[h]
\caption{Parameters for Short-Time Fourier Transform (STFT) and mel-spectrogram extraction given time-domain speech waveforms.}
\label{stft_param}
\begin{center}
\begin{small}
\setlength{\tabcolsep}{5pt}
\begin{tabular}{ll}
\toprule
\textbf{Parameter} & \textbf{Value} \\
\midrule
Sample Rate & 24 kHz \\
Minimum Frequency & 0 Hz \\
Maximum Frequency & 12 kHz \\
Mel Bins & 80 \\
FFT Size & 2048 \\
Hop Size & 240 \\
Window Size & 1200 \\
Signal Normalization & True \\
Symmetric Mels & True \\
Maximum Absolute Value & 4.0 \\
Minimum Decibel Level & -115.0 \\
\bottomrule
\end{tabular}
\end{small}
\end{center}
\vskip -0.1in
\end{table*}

\begin{table*}[b]
\caption{Models used from \href{https://learn.microsoft.com/en-us/azure/ai-services/speech-service/language-support?tabs=tts}{Microsoft Azure AI platform} to generate parallel accented speech for BN2BN training.}
\label{tts_models}
\begin{center}
\begin{small}
\setlength{\tabcolsep}{5pt}
\begin{tabular}{cll}
\toprule
\textbf{Language} & \textbf{Accent} & \textbf{Model} \\
\midrule
\multirow{12}{*}{\shortstack{English}} & American & en-US-GuyNeural \\
 & Australian & en-NZ-MitchellNeural \\
 & British & en-GB-RyanNeural \\
 & Chinese & zh-CN-XiaohanNeural \\
 & French & fr-FR-BrigitteNeural \\
 & Greek & el-GR-AthinaNeural \\
 & Indian & en-IN-NeerjaNeural \\
 & Italian & it-IT-DiegoNeural \\
 & Kenyan & en-KE-AsiliaNeural \\
 & Korean & ko-KR-SoonBokNeural \\
 & Mexican & es-MX-CandelaNeural \\
 & Russian & ru-RU-DmitryNeural \\
\hline
 \multirow{8}{*}{\shortstack{Mandarin\\Chinese}} & Guangxi & zh-CN-guangxi-YunqiNeural \\
 & Henan & zh-CN-henan-YundengNeural \\
 & Jilu & zh-CN-shandong-YunxiangNeural \\
 & Liaoning & zh-CN-liaoning-XiaobeiNeural \\
 & Mandarin & zh-CN-YunxiNeural  \\
 & Shaanxi & zh-CN-shaanxi-XiaoniNeural \\
 & Sichuan & In-house dataset of human speaker \\
 & Taiwanese & zh-TW-HsiaoChenNeural  \\
 \bottomrule
\end{tabular}
\end{small}
\end{center}
\vskip -0.1in
\end{table*}

\begin{table*}[t]
\caption{Detailed accent-wise results of monolingual and cross-lingual accent conversion objective evaluation, measuring conversion strength. For each target accent, we provide the averaged cosine similarity of embeddings extracted from accent classifiers and denote the number of individual samples used to calculate each value in parentheses.}
\label{accent_cosine_similarity_by_accent}
\begin{center}
\begin{small}
\setlength{\tabcolsep}{5pt}
\begin{tabular}{clccc}
\toprule
\textbf{Conversion Type} & \textbf{Target Accent} & \textbf{Ground Truth} & \textbf{Model Output} & \textbf{Non-Matching Accent}\\
\midrule
\multirow{11}{*}{\shortstack{English\\Monolingual}} & To American & 0.997$\pm$0.001 (100) & 0.779$\pm$0.054 (100) & 0.039$\pm$0.017 (90) \\
& To Australian & 0.990$\pm$0.010 (100) & 0.702$\pm$0.072 (100) & 0.033$\pm$0.009 (90) \\
& To Chinese & 0.998$\pm$0.001 (100) & 0.878$\pm$0.029 (100) & 0.034$\pm$0.013 (90) \\
& To French & 0.998$\pm$0.001 (100) & 0.829$\pm$0.042 (100) & 0.044$\pm$0.016 (90) \\
& To Greek & 0.986$\pm$0.003 (100) & 0.805$\pm$0.044 (100) & 0.033$\pm$0.011 (90) \\
& To Indian & 0.998$\pm$0.001 (100) & 0.696$\pm$0.061 (100) & 0.018$\pm$0.014 (90) \\
& To Italian & 0.998$\pm$0.001 (100) & 0.931$\pm$0.024 (100) & 0.004$\pm$0.011 (90) \\
& To Kenyan & 0.998$\pm$0.001 (100) & 0.838$\pm$0.046 (100) & -0.026$\pm$0.010 (90) \\
& To Mexican & 0.999$\pm$0.001 (100) & 0.840$\pm$0.041 (100) & -0.025$\pm$0.013 (90) \\
& To Russian & 0.996$\pm$0.001 (100) & 0.677$\pm$0.060 (100) & 0.025$\pm$0.012 (90) \\
\cmidrule{2-5}
& Overall & 0.996$\pm$0.001 (1,000) & \textbf{0.798$\pm$0.016} (1,000) & 0.018$\pm$0.004 (900) \\
\midrule
\multirow{9}{*}{\shortstack{Mandarin\\Monolingual}} & To Guangxi & 0.998$\pm$0.001 (64) & 0.981$\pm$0.003 (64) & 0.689$\pm$0.055 (56) \\
& To Henan & 0.989$\pm$0.007 (64) & 0.931$\pm$0.022 (64) & 0.612$\pm$0.031 (56) \\
& To Jilu & 0.988$\pm$0.010 (64) & 0.867$\pm$0.060 (64) & 0.519$\pm$0.056 (56) \\
& To Liaoning & 0.989$\pm$0.003 (64) & 0.929$\pm$0.009 (64) & 0.649$\pm$0.058 (56) \\
& To Mandarin & 0.954$\pm$0.021 (64) & 0.840$\pm$0.036 (64) & 0.653$\pm$0.058 (56) \\
& To Shaanxi & 0.992$\pm$0.002 (64) & 0.910$\pm$0.025 (64) & 0.630$\pm$0.043 (56) \\
& To Sichuan & 0.994$\pm$0.001 (64) &  0.967$\pm$0.006 (64) & 0.582$\pm$0.051 (56) \\
& To Taiwanese & 0.994$\pm$0.001 (64) &  0.892$\pm$0.030 (64) & 0.636$\pm$0.043 (56) \\
\cmidrule{2-5}
& Overall & 0.987$\pm$0.003 (512) & \textbf{0.915$\pm$0.011} (512) & 0.621$\pm$0.018 (448) \\
\midrule
\multirow{9}{*}{\shortstack{Cross-Lingual}} & To Mandarin & 0.940$\pm$0.013 (64) & 0.758$\pm$0.035 (64) & 0.549$\pm$0.032 (56) \\
& To Shaanxi & 0.988$\pm$0.003 (64) & 0.891$\pm$0.020 (64) & 0.629$\pm$0.046 (56) \\
& To Sichuan & 0.995$\pm$0.001 (64) & 0.972$\pm$0.003 (64) & 0.668$\pm$0.047 (56) \\
& To Taiwanese & 0.990$\pm$0.002 (64) & 0.891$\pm$0.018 (64) & 0.494$\pm$0.039 (56) \\
& To American & 0.988$\pm$0.003 (64) & 0.808$\pm$0.046 (64) & 0.588$\pm$0.055 (56) \\
& To French & 0.993$\pm$0.002 (64) & 0.772$\pm$0.040 (64) & 0.435$\pm$0.049 (56) \\
& To Indian & 0.989$\pm$0.002 (64) & 0.661$\pm$0.060 (64) & 0.650$\pm$0.052 (56) \\
& To Kenyan & 0.992$\pm$0.002 (64) & 0.871$\pm$0.034 (64) & 0.390$\pm$0.042 (56) \\
\cmidrule{2-5}
& Overall & 0.984$\pm$0.002 (512) & \textbf{0.828$\pm$0.015} (512) & 0.550$\pm$0.018 (448) \\
\bottomrule
\end{tabular}
\end{small}
\end{center}
\end{table*}

\end{document}